\documentclass[a4paper,journal,twocolumn]{IEEEtran}
\usepackage{amsmath,amsfonts}
\usepackage{algorithmic}
\usepackage{algorithm}
\usepackage{array}
\usepackage{color}
\usepackage{subfig}
\usepackage{bm}
\usepackage{textcomp}
\usepackage{float}
\usepackage{subfig}
\usepackage{url}
\usepackage{verbatim}
\usepackage{graphicx}
\usepackage{cite}
\usepackage{geometry}
\usepackage[table,svgnames]{xcolor}
\usepackage[algo2e,ruled,vlined]{algorithm2e}
\usepackage{array}
\usepackage{epsfig}
\newcommand{\RNum}[1]{\uppercase\expandafter{\romannumeral #1\relax}}

\begin{document}
\title{RWZC: A Model-Driven Approach for Learning-based Robust Wyner-Ziv Coding}

\author{Yuxuan Shi, 
Shuo Shao, \IEEEmembership{Member,~IEEE}, 
Yongpeng Wu, \IEEEmembership{Senior Member,~IEEE}, 
Wenjun Zhang, \IEEEmembership{Fellow,~IEEE},
and Merouane Debbah \IEEEmembership{Fellow,~IEEE}
\vspace{-0.3cm}
\thanks{(corresponding authors: Shuo Shao, Yongpeng Wu)

Yuxuan Shi is with the department of networked intelligence, Pengcheng Laboratory, Shenzhen 410083, China(e-mail: shiyx01@pcl.ac.cn), was with the School of Cyber and Engineering, Shanghai Jiao Tong University, Shanghai 200240, China (e-mail: ge49fuy@sjtu.edu.cn) 

Shuo Shao is with the School of Cyber and Engineering, Shanghai Jiao Tong university, Shanghai 200240, China (e-mail: shuoshao@sjtu.edu.cn) 

Yongpeng Wu and Wenjun Zhang is with the Department of Electronic Engineering, Shanghai Jiao Tong University, Shanghai 200240, China (e-mail: yongpeng.wu@sjtu.edu.cn, zhangwenjun@sjtu.edu.cn).

M. Debbah is with KU 6G Research Center, Khalifa University of Science and Technology, P O Box 127788, Abu Dhabi, UAE (email: merouane.debbah@ku.ac.ae) and also with CentraleSupelec, University Paris-Saclay, 91192 Gif-sur-Yvette, France.
}
}
%\markboth{}%


\maketitle
\pagestyle{empty}
\thispagestyle{empty}
\begin{abstract}
		In this paper, a novel learning-based Wyner-Ziv coding framework is considered under a distributed image transmission scenario, where the correlated source is only available at the receiver. Unlike other learnable frameworks, our approach demonstrates robustness to non-stationary source correlation, where the overlapping information between image pairs varies. Specifically, we first model the affine relationship between correlated images and leverage this model for learnable mask generation and rate-adaptive joint source-channel coding. Moreover, we also provide a warping-prediction network to remove the distortion from channel interference and affine transform. Intuitively, the observed performance improvement is largely due to focusing on the simple geometric relationship, rather than the complex joint distribution between the sources. Numerical results show that our framework achieves a $1.5$ dB gain in PSNR and a $0.2$ improvement in MS-SSIM, along with a significant superiority in perceptual metric, compared to state-of-the-art methods when applied to real-world samples with non-stationary correlations.
	\end{abstract}
	

	\section{Introduction}
	\IEEEPARstart{T}{he} rapid progress in mobile devices and wireless networks has sparked growing interest in multi-node communications, particularly in applications such as autonomous driving assistance, virtual reality, multi-camera surveillance, and medical imaging. A key challenge in these scenarios is the need for edge devices to transmit data at low rates while maintaining reliability and minimizing latency. Distributed source coding (DSC) offers a promising solution by exploiting the statistical dependencies between correlated sources, allowing them to be encoded without direct inter-source communication. This approach improves encoding efficiency and reduces data redundancy.
	
	Research on DSC dates back to the 1970s, when Slepian and Wolf \cite{SlepianW73} established the optimal rate region for independently encoding two correlated sources (SW problem). This was followed by the study of the Wyner-Ziv (WZ) problem (source coding with decoder-only side information) \cite{WynerZ76}, the multi-terminal (MT) problem (source coding for correlated multiple users) \cite{Tung78}, and the chief executive officer (CEO) problem (source coding for correlated sources with a common remote source) \cite{BergerZV96}. These foundamental works provide key theoretical insights that have guided the design of compression and transmission systems in real-world applications. When combining DSC with transmission, the integration of source and channel coding becomes especially critical. Joint source-channel coding (JSCC) has been shown to outperform separate source-channel coding in scenarios with limited block lengths, despite its higher implementation complexity. However, this problem has been solved by the rapid advances in deep learning (DL) technologies. In recent years, numerous convolutional neural network (CNN) or Transformer-based frameworks for distributed compression and transmission have emerged, particularly for image sources \cite{WangYDN22, WangLMF22, YilmazKG23, AyzikA20, MitalOGG22, MitalOGG23, YilmazKG23_2, BritesAP12, ZhangT21}. For scenarios where both correlated sources require encoding (SW problem), Wang et al. \cite{WangYDN22} proposed a deep neural network solution incorporating a cross-attention module being aware of channel state information. In \cite{WangLMF22}, the authors explored distributed audio-visual parsing using a multi-modal Transformer and DL-based JSCC. In the case where side information is only available at the decoder (WZ problem), Ayzik and Avidan \cite{AyzikA20} introduced a side-information-finder block to perform plane-wise projection for DL-based image compression, termed the Deep Side Information Network (DSIN). Mital et al. \cite{MitalOGG22, MitalOGG23} further improved Ayzik's work by integrating an entropy model and cross-attention mechanism to create a more efficient compression scheme, known as Neural Distributed Image Compression (NDIC). More recently, Yilmaz et al. \cite{YilmazKG23_2} proposed a novel framework, DeepJSCC-WZ, which integrates an attention feature module and achieves low-latency joint source-channel coding for transmission scenarios.
	
\begin{figure}[tbp]
	\centering
	\vspace{0.15cm}
	\begin{minipage}[t]{0.3\linewidth}
		\centering
		\includegraphics[width=1\textwidth]{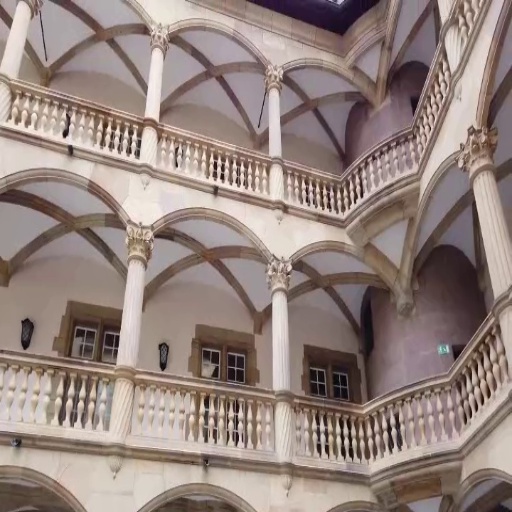}
		\vspace*{0.001cm} 
	\end{minipage}
	\begin{minipage}[t]{0.3\linewidth}
		\centering
		\includegraphics[width=1\textwidth]{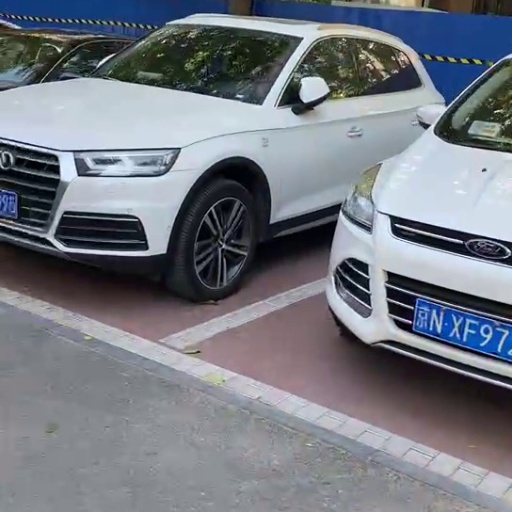}
	\end{minipage}
	\begin{minipage}[t]{0.3\linewidth}
		\centering
		\includegraphics[width=1\textwidth]{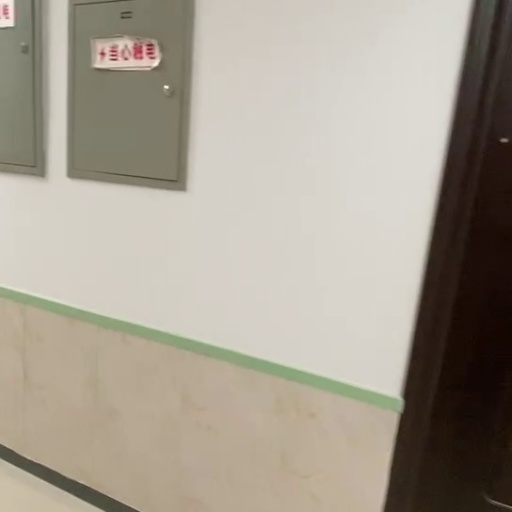}
	\end{minipage}
	\begin{minipage}[t]{0.3\linewidth}
		\centering
		\includegraphics[width=1\textwidth]{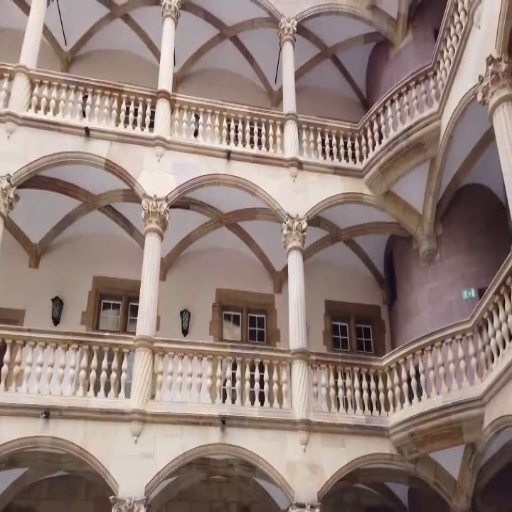}
		%\vspace*{1ex} 
		\begin{center}
			\small high correlation
		\end{center}
	\end{minipage}
	\begin{minipage}[t]{0.3\linewidth}
		\centering
		\includegraphics[width=1\textwidth]{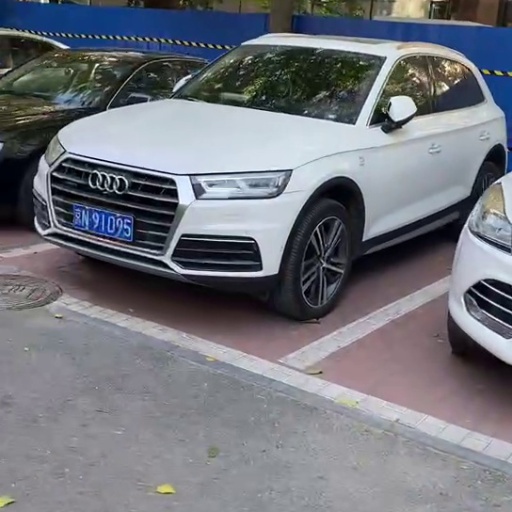}
		\begin{center}
			\small medium correlation
		\end{center}
	\end{minipage}
	\begin{minipage}[t]{0.3\linewidth}
		\centering
		\includegraphics[width=1\textwidth]{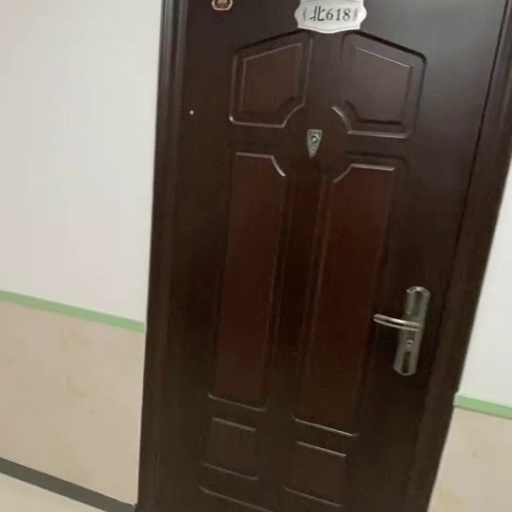}
		\begin{center}
			\small	low correlation
		\end{center}
	\end{minipage}
	\captionsetup{font={small}}
	\caption{Illustration for image pairs of different correlation/parallax (from dataset UDIS-D), where parallax means the visual deviation between two images caused by different positions}\label{illu}
\end{figure}

	However, the existing frameworks for both SW and WZ problems primarily focus on highly correlated and stably distributed sources to achieve optimal compression or transmission performance, which does not accurately reflect the variability of real-world data. In scenarios where source correlation fluctuates significantly, such as image pairs with varying parallax (as shown in Fig. \ref{illu}), these methods often fail to achieve satisfactory results. This performance gap can be attributed to three main reasons. First, when training on images with non-stationary correlations, instability arises during feature extraction, leading to mismatches and inaccurate representations. Second, CNN-based architectures often have insufficient receptive fields to capture the correlations in highly uncorrelated images, resulting in confusion among convolutional features and a decline in performance. Third, the non-stationary distribution of correlation in the input data significantly reduce the network's ability to generalize across different data samples. In summary, the performance of feature-level interaction heavily relies on the stability of input distribution in data-driven architectures, thus it performs unsatisfactorily in situations where the correlation distribution fluctuates, inevitably. In such cases, a well-designed correlation model can effectively address these issues.

%	the modeling the correlation between images in the WZ problem, where we call the correlation as source side information (SSI). A common solution based on the perspective transform is to use a linear homography model, which is typically represented by a $3\times3$ matrix. By establishing a high-quality homography relationship, we can decouple the correlation between the transmitted image and the side image, which avoids the interaction of this correlation at feature level and allows us to transmit a decoupled portion of the original image at a low rate. Meanwhile, at the decoder side, this portion is combined with the warped side image to complete the reconstruction
	Therefore, we assume that there exists a conception, named as source state information (abbreviated by SSI), which used to describe the correlation between sources with low complexity. This approach facilitates source decorrelation, thereby preventing mismatched representations and feature confusion. To validate the effectiveness of this conception, we consider a practical WZ scenario in this paper involving camera arrays, where two stereo cameras capture images of the same object. One camera transmits photos $\bm{x}$ to the receiver, while the receiver has access to the noise-free image $\bm{y}$ from the other camera. Note that, we assume the correlation between two images is non-stationary owing to the changeable camera poses, camera parameters or even a movable objects. To address the performance degradation caused by varying correlation, we propose three key components driven by the modeling of SSI: correlation estimation and decoupling, low-rate transmission, and SSI-aided source reconstruction. Specifically, we first estimate the SSI between the image pairs and use it for source decoupling of $\bm{x}$, where "decoupling" refers to removing the redundant information carried by the side image $\bm{y}$. The SSI is also available at the decoder without error. Next, we apply an efficient JSCC scheme to transmit the decoupled image with limited resources. Finally, $\bm{y}$, along with SSI, aids in the reconstruction of $\bm{x}$ at the decoder.
	
	Following this idea, we propose a novel learning-based  joint source-channel coding framework for images, named as Robust Wyner-Ziv coding (abbreviated by RWZC), which ensures reliable transmission of sources with non-stationary correlation. The key contributions of this paper are summarized as follows:
	\begin{itemize}
		\item We propose a perspective transform layer (PTL)-based mask generation module for source decoupling. This module learns the dynamic distribution of the SSI and generates adaptive masks to decorrelate the transmitted image from the side image. This approach overcomes the issues of corrupted masking encountered in traditional methods. Additionally, since the decoupled image is significantly smaller than the original, RWZC offers superior performance in low-rate transmission compared to other schemes.
		\item We propose a CNN-based autoencoder that implements low-rate JSCC, encoding the decoupled image into feature representations whose size adapts to the masked regions. This rate-adaptive mechanism utilizes non-linear transform (NLT) and entropy modeling of each feature component, effectively saving the transmission resources.
		\item We introduce a warping prediction reconstruction network designed to reduce distortions caused by channel interference and perspective transform. An SSI-integrated mechanism is incorporated for mask prediction and distortion compensation, helping to repair pixel-level corruption in image reconstruction due to noise interference and resolution discrepancies.
\end{itemize}
Simulation results demonstrate that our framework offers advantages over state-of-the-art frameworks in the WZ scenario, tested on datasets with both stationary and non-stationary correlated image pairs. First, for stationary correlated samples, RWZC performs similarly to the baseline methods on reconstruction metrics. However, on the non-stationary dataset, it outperforms the others due to its affine correlation modeling and efficient transmission. Second, our framework achieves a substantial advantage at the perceptual scale, as slight pixel shifts have minimal impact on human perception. Finally, cross-dataset testing validates the high robustness of RWZC when handling non-stationary distributed sources.

The paper is organized as follows: in Section \ref{section:system model}, we first introduce the decoder-only side information assistant model for image transmission, and simply introduce the RWZC framework. In Section \ref{Sec3}, the detailed approach is presented, including the homography estimation, low rate JSCC encoding and SSI-aided reconstruction. Furthermore, the numerical results are shown in Section \ref{Sec5}, illustrating the superiority of our frameworks. Finally, Section \ref{Sec6} concludes the paper.

% \section{Metaverse and Semantic Communication}
	\begin{figure*}[htbp]
	\centering
	\includegraphics[width=0.98\textwidth]{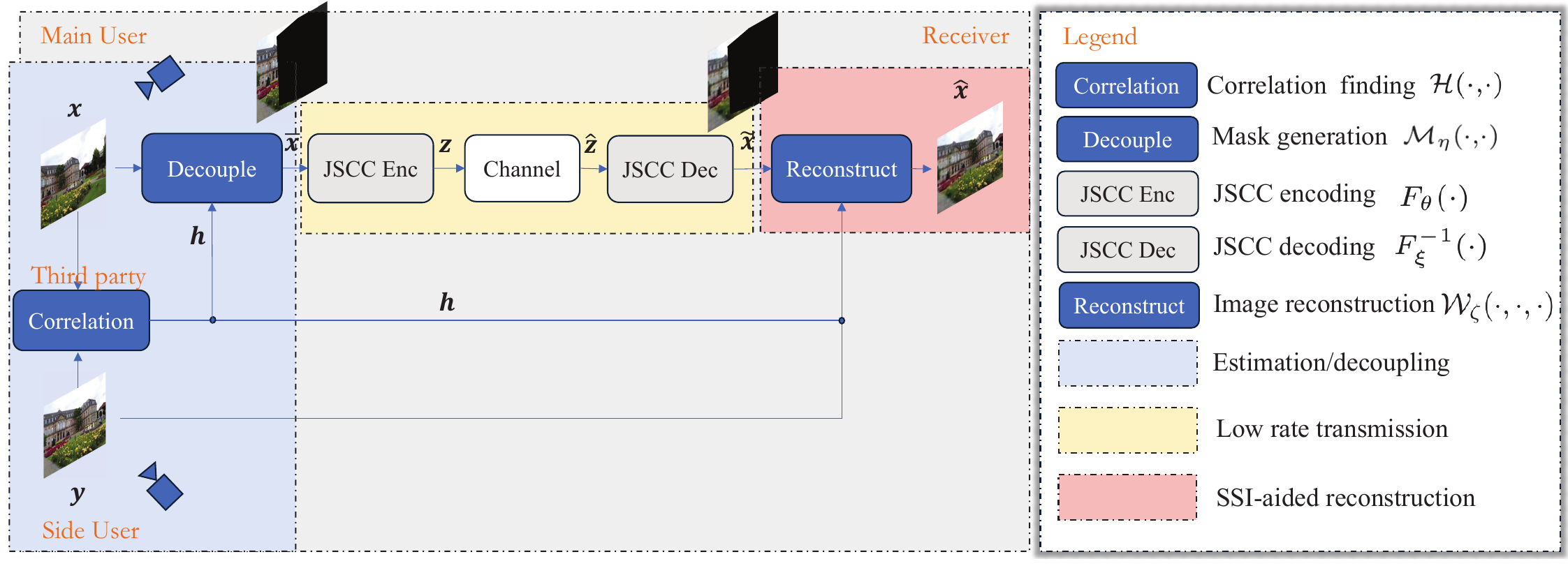}
	\captionsetup{font={small},justification=raggedright}
	\caption{An overview of the RWZC framework.}
	\label{sys}
\end{figure*}
	\vspace{-0.1cm}
	\section{System Model and Proposed Framework}
	\label{section:system model}
	In this section, we first introduce the notations and the system model for the Wyner-Ziv coding scenario applied to image transmission. Additionally, we provide a generalized overview of RWZC, which includes correlation estimation, low-rate transmission, and SSI-aided image reconstruction.

\vspace{-0.2cm}
	\subsection{Notation Convenience}
	Throughout this paper, the realization of the corresponding random variable is denoted by a lowercase letter, while a scalar is represented in normal font and a vector in bold font. For instance, we utilize the scalar $x$ and vector $\bm{x}=(x_1,x_2,\cdots,x_L)$. Moreover, we denote the sets of real numbers and complex numbers $\mathbb{R}$ and $\mathbb{C}$, respectively. Besides, $\bm{z}^H$ denotes the Hermitian transpose of matrix $\bm{z}$.
\vspace{-0.1cm}
	\subsection{System Model}
	We consider an image transmission scenario using two stereo cameras, where the cameras are positioned at different angles to capture images of the same object, resulting in images with overlapping information. It is assumed that the camera positions are fixed. In this setup, one set of images is transmitted over a wireless channel, while another set of images, transmitted over a high-rate/error-free channel, is used at the receiver to assist the reconstruction of the first set. This setup, where only side information is available at the receiver, is known as the WZ coding problem. Specifically, the main camera captures image $\bm{x}\in\mathbb{R}^{H\times W\times C}$ and encodes it into a $k$-length codeword $\bm{z}\in\mathbb{R}^k$, subject to an average power constraint: $\frac{1}{k}||\bm{z}||^2 \leq P_{\mathrm{ow}}$. The codeword is then transmitted through an additive white Gaussian noise (AWGN) channel, which is modeled as
	\begin{align}
		\hat{\bm{z}}=\bm{z}+\bm{n}\label{channel}
	\end{align}
 where $\bm{n}\sim\mathcal{CN}(\bm{0},\sigma_n^2\bm{I}_k)$ denotes the unbiased complex Gaussian noise. For further use, the metric named channel bandwidth ratio (CBR) is defined as $\frac{k}{H\times W\times C}$ to measure the resources per pixel consumed for the transmission, and channel signal-to-noise ratio (SNR) is defined as $10\log_{10}(\frac{P_{\mathrm{ow}}}{\sigma_n^2})$ with unit dB. Finally, the decoder utilizes the noisy codewords to recover $\bm{x}$ with the help of side image $\bm{y}\in\mathbb{R}^{H\times W\times C}$ from the secondary camera, and finally outputs the reconstruction $\hat{\bm{x}}$. 

 The core difference from other frameworks is, we utilize a conception to describe the correlation between sources. The word "source state information", whose naming inspired by channel state information (CSI), refers to the specific relation between the sources $(\bm{x},\bm{y})$, and is available at both encoder and decoder without errors, as illustrated in Fig. \ref{sys}.  In our framework, the availability of SSI at the encoder does not chanllenge the setting of WZ coding, which is mainly due to the following fact. In general, SSI can be interpreted as \textbf{a parameter of the joint distribution} $P_{\bm{XY}}$ and estimated at a third party. Even when the encoder acquires the parameter of the source joint distribution, it remains incapable of directly using the side information, which does not violate the criterion that $\bm{x}$ and $\bm{y}$ cannot interact at the encoder in WZ problem. Note that SSI in RWZC functions similarly to CSI in wireless transmission, where the attainment of CSI does not necessarily mean the acquisition of exact channel noise \cite{DBLP:journals/tnse/LuoJWCL20,DBLP:journals/tcom/GuoWJL22}.

 In fact, as a dual case, considering the implicit test channel $P_{\bm{Y|X}}$ existing between correlated sources, the connection between SSI and CSI goes beyond this. For instance, both of them can be estimated apart from the encoder and fed back to optimize the transmission strategy. Besides, SSI and CSI describe source correlation and channel state distributions, respectively, but independent of the information to be compressed/transmitted. Similarly, we can draw an analogy between sources with stationary correlation distributions and block fading channels, while another analogy between sources with non-stationary correlation distributions and fast fading channels. Overall, like CSI, SSI is a characterization of system parameters, and its use in the encoder does not affect the settings of the original problem.
 \vspace{-0.2cm}
\subsection{Proposed framework of RWZC}
	As a target, RWZC is expected to realize a bandwidth-friendly and reliable transmission for non-stationary correlated images at a WZ scenario, via the assistance of SSI. Specifically, RWZC consists of three main components. First, correlation estimation evaluates the SSI between image pairs using a pre-established module that detects affine transform, and sends this correlation to both the transmitter and receiver without errors. This allows the transmitter to decorrelate $\bm{x}$ from $\bm{y}$. Second, a joint source-channel (JSC) codec encodes and transmits the decoupled image over the channel, ensuring efficient and reliable transmission. Third, a reconstruction module at the receiver utilizes the correlation to achieve high-quality recovery of the transmitted $\bm{x}$. The detailed components are outlined as follows:
	\subsubsection{\textbf{Correlation estimation and decoupling}} 
At the first step, RWZC estimates the relation between image pairs and decouples $\bm{x}$ from side image $\bm{y}$. In our framework, the interaction between $(\bm{x},\bm{y})$ occurs at a third party via a predefined model $\mathcal{H}(\cdot,\cdot)$, which outputs the SSI between two images as $\bm{h}$. Note that a reasonable modeling significantly affects the performance of RWZC, while in the latter we will introduce a widely-used geometric method describing correlated images of camera arrays, named homography, which estimates the affine relation among images with low complexity. After that, by delivering $\bm{h}$ to the decoupling model at the transmitter, image $\bm{x}$ can be separated into portions irrelevant and relevant to the side image $\bm{y}$, which is achieved using a learning-based masking strategy. The overall estimating and decoupling process can be formulated as
\vspace{-0.26cm}
\begin{align}	
(\bm{x},\bm{y})&\xrightarrow{\mathcal{H}(\cdot,\cdot)}\bm{h},\\
(\bm{x},\bm{h})&\xrightarrow{\mathcal{M}_\eta(\cdot,\cdot)}\bar{\bm{x}}\in\mathbb{R}^{H\times W\times C},\label{3}
\end{align}
where $\eta$ represents the learnable weights of the network. Note that in RWZC, the encoder can only access SSI for optimzing transmission strategy, as stated in Eq. \eqref{3}. 
%Note that the delivery is only necessary at the training phase for learning a distribution of SSI (marked as dashed arrow in Fig. \ref{sys}), since the framework can adaptively select a proper decorelation strategy according to different $\bm{h}$ when inferencing.
	\subsubsection{\textbf{Low rate transmission}}
At the second part, RWZC conducts a bandwidth-friendly transmission based on a learnable joint source-channel codec. Now equipped with a decoupled image $\bar{\bm{x}}$ which is a portion of the original image, the JSC encoder conducts feature extraction and transmission of $\bar{\bm{x}}$ for a reliable and low rate transmission through a channel. The encoding process at the transmitter can be formulated as\footnote{Herein we use the notation $F_\theta$ to represent the learnable function, while subscript $\theta$ denotes the weights. Functions $F^{-1}_\xi$ and $\mathcal{W}_\zeta$ follow similarly.}
\begin{align}
	\bar{\bm{x}}&\xrightarrow{F_\theta(\cdot)}\tilde{\bm{z}},
\end{align}
where $\tilde{\bm{z}}$ denotes the latent vector and the normalized outputs $\bm{z}=\sqrt{kP_{\mathrm{ow}}}\frac{\tilde{\bm{z}}}{\sqrt{\tilde{\bm{z}}^H\tilde{\bm{z}}}}$. It should be noticed that the decoupled image $\bar{\bm{x}}$ has different sizes, since the side image $\bm{y}$ contains different overlapping content in each image pair. This hence results in a bandwidth waste if using a fixed-length coding strategy, or a full connected (FC) layer of fixed output dimensions in neural networks. In the later we will introduce a rate-adaptive JSC framework for efficient transmission based on the non-linear transform, in which each component of extracted features can be adapted to its corresponding entropy and thus improve the transmission efficiency.  
%Besides, 
\subsubsection{\textbf{SSI-aided source reconstruction}}
At the final step, our framework completes the reconstruction of the original image $\bm{x}$ utilizing SSI and the side image $\bm{y}$. Specifically, after the noisy transmission defined in Eq. \eqref{channel}, we obtain distorted features of the decoupled image $\hat{\bm{z}}\in\mathbb{R}^k$ at the receiver, which is then recovered to noisy decoupled image $\tilde{\bm{x}}$ by the JSC decoder. Meanwhile, the receiver is allowed to access the side image $\bm{y}$ and SSI $\bm{h}$, thus reconstructs $\hat{\bm{x}}$ via a learnable network $\mathcal{W}_\zeta(\cdot,\cdot,\cdot)$. The total process is 
\begin{align}
	\hat{\bm{z}}&\xrightarrow{F^{-1}_\xi(\cdot)}\tilde{\bm{x}}\in\mathbb{R}^{H\times W\times C},\\
	(\tilde{\bm{x}},\bm{y},\bm{h})&\xrightarrow{\mathcal{W}_\zeta(\cdot,\cdot,\cdot)}\hat{\bm{x}}\in\mathbb{R}^{H\times W\times C},
\end{align}
where we use the decoding function $F^{-1}_\xi(\cdot)$ to represent the inverse operation of encoding. 

The three main components of RWZC have been introduced above. In addition, $\mathcal{H}(\cdot,\cdot)$, which estimates the correlation between image pairs, is a non-learnable module, while the remaining components are all learning-based. In the next section, \textbf{we reveal the fact that RWZC framework focuses on estimating simple geometric parameters between $(\bm{x}, \bm{y})$ rather than directly evaluating the complex joint distribution $P_{\bm{XY}}$.} This approach provides better generalization to sources with varying correlations and significantly reduces complexity.
%\vspace{-0.2cm}
\section{Detailed Methodologies}\label{Sec3}
This section presents the detailed schemes to realize the aforementioned modules. Specifically, for correlation estimation, we utilize the geometric method named as homography, which can be evaluated by existing feature matching algorithms. For decoupling method, an SSI-robust masking strategy is proposed for varying $\bm{h}$ between different image pairs, which is realized by a learnable perspective transform. Moreover, for transmission, a CNN-based autoencoder with hyper-prior is adopted for rate adaptive transmission. Finally, for high-quality reconstruction, a distortion prediction module is proposed at the receiver for accurate warping of side image and seamless blending.
\vspace{-0.4cm}
\subsection{Homography estimation}\label{subsecSIFT}
\begin{figure}[tbp]
	\centering
	\includegraphics[width=0.45\textwidth]{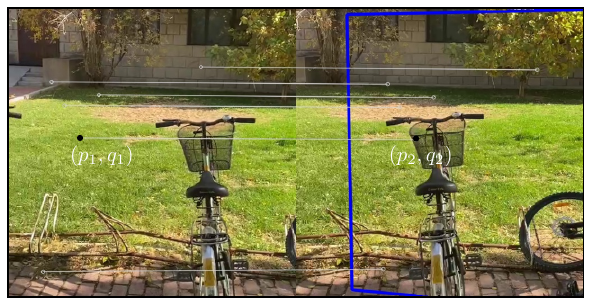}
	\captionsetup{font={small}}
	\caption{Feature matching based homography estimation, in which $(p_1,q_1)$ and $(p_2,q_2)$ are the matching points pair}
	\label{SIFT}
\end{figure}

For our correlation estimation module $\mathcal{H}$, we adopt an perspective transform (PT)-based method named as homography for correlation modeling, which describes the projective relation from 3D world to 2D plane, as illustrated in Fig. \ref{SIFT}.

A homography matrix is utilized for modeling the affine correlation between image pairs, which is a transform matrix used to map points from one plane to another in projective geometry. In computer vision and image processing, homography matrices are commonly employed to represent the transform between two images of the same scene taken from different viewpoints or perspectives. Note that homography always appears as a matrix of size $3\times3$ with only 8 degrees of freedom. The extreme low complexity enables the negligible cost for error-free transmission of $\bm{h}$ to encoder and decoder, which motivates us to name it as source state information.

For the homography estimation \cite{0015576,Zhang00}, given an image pair, we assume there exists a matrix yielding the affine transform between two images from one point $(p^i_1,q^i_1)$ to its corresponding point $(p^i_2,q^i_2)$, where $i=1,2,\cdots,S$ and $S$ denotes the number of all matching points, namely
\begin{align}
	\begin{bmatrix}
		p^i_2\\
		q^i_2\\
		1
	\end{bmatrix}=
\bm{h}\begin{bmatrix}
	p^i_1\\
	q^i_1\\
	1
\end{bmatrix}=
	\begin{bmatrix}
		h_{11}&h_{12}&h_{13}\\
		h_{21}&h_{22}&h_{23}\\
		h_{31}&h_{32}&h_{33}
	\end{bmatrix}
	\begin{bmatrix}
		p^i_1\\
		q^i_1\\
		1
	\end{bmatrix}.
\end{align}
Consequently the corresponding point pair $(p_2,q_2)$ is formulated as
\begin{align}
	p^i_2=\frac{h_{11}p^i_1+h_{12}q^i_1+h_{13}}{h_{31}p^i_1+h_{32}q^i_1+h_{33}};
	q^i_2=\frac{h_{21}p^i_1+h_{22}q^i_1+h_{23}}{h_{31}p^i_1+h_{32}q^i_1+h_{33}}
\end{align}
 and thus each pair of points contributes two equations for solving the homography as shown in the following.
 \begin{align}
 	\begin{bmatrix}
 		p^i_1 &q^i_1 &1 &0 &0 &0 &-p^i_2p^i_1& -p^i_2p^i_1& -p^i_2\\
 		0 &0 &0 &p^i_1 &q^i_1 &1 &-q^i_2p^i_1& -q^i_2q^i_1& -q^i_2
 	\end{bmatrix}\tilde{\bm{h}}=\begin{bmatrix}
 0\\
 	0
 \end{bmatrix},\notag
 \end{align}
where $\tilde{\bm{h}}=[h_{11},h_{12},h_{13},h_{21},\cdots,h_{33}]$. This means that the homography estimation only needs four non-collinear coordinates pairs due to the 8 free parameters in $\bm{h}$. After that, this overdetermined system will be solved with the best four geometric-related pairs, which is found by algorithms. In a word, this pixel-wised affine connection uniquely determine the overlapping region between image pair.

The overall homography estimation process can be summarized in the following five steps. First, feature points of each image pair are extracted by conventional algorithm, e.g. SIFT/SURF/FAST/ORB \cite{RubleeRKB11,Lowe04,BayTG06,RostenD06}. Meanwhile, the corresponding descriptors are also evaluated, which is used to describe the local features. After that, a proper matcher is adopted for feature matching between image pairs with a pre-defined threshold, like Brute-Force, FLANN, Optical Flow \cite{MujaL09,HornS81}. Moreover, Random Sample Consensus (RANSAC) \cite{FischlerB81} estimates homography parameters by randomly selecting sample points from the matching points, and calculating the goodness of model fitting. At this step, defected points will be excluded and 8 equations will be solved, and finally outputs the estimated homography. 

We note that $\bm{h}$ represents the parameters in the affine model of source correlation, capturing the relative positional information between images. In fact, $\bm{h}$ is determined only by the capture angles of the images, and is independent of the images content, yielding the inequivalence between $\bm{h}$ and $\bm{y}$. This further supports the claim that RWZC utilizes $\bm{h}$ without modifying the WZ setups.
%Note that the performance of the traditional algorithm highly depends on the features extraction and defects exclusion.
\vspace{-0.3cm}
\subsection{Homography-adaptive masking strategy}\label{decoupling}
\begin{figure}[tbp]
	\centering
	\includegraphics[width=0.46\textwidth]{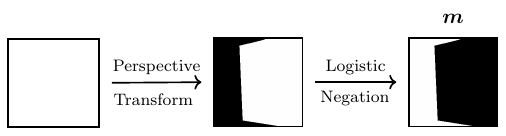}
	\captionsetup{font={small},justification=raggedright}
	\caption{Mask generation via homography mapping.}
	\label{mask}
\end{figure}

\begin{figure}[bp]
	\centering
	\vspace{-0.25cm}
	\subfloat[original image]{
		\includegraphics[width=0.15\textwidth]{figures/pic/000744.jpg}}
	\subfloat[side image]{
		\includegraphics[width=0.15\textwidth]{figures/pic/000744right.jpg}}
	\subfloat[A fail recovery]{
		\includegraphics[width=0.151\textwidth]{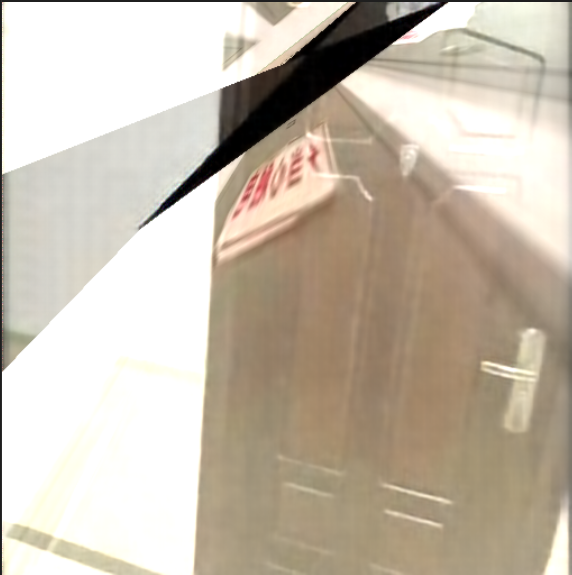}}
	\newline
	\centering
	\subfloat[original image]{
		\includegraphics[width=0.15\textwidth]{figures/pic/000518.jpg}}
	\subfloat[side image]{
		\includegraphics[width=0.15\textwidth]{figures/pic/000518right.jpg}}
	\subfloat[A fail recovery]{
		\includegraphics[width=0.151\textwidth]{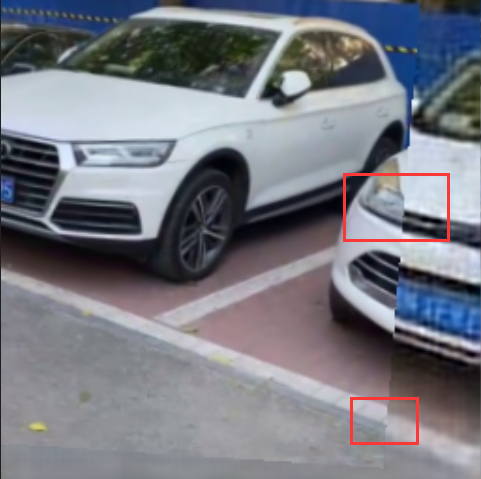}}
	\captionsetup{font={small}}
	\caption{Fail cases of non-learnable masking on image pairs with: large parallax (First row);  varying parallax (Second row)}\label{Fig_fail}
\end{figure}
Considering the image decoupling module, we utilize a masking strategy for removing the redundant information which has been provided by side image $\bm{y}$ at the decoder. To do this, we generate a transform-based mask for each $\bm{x}$ to remove the unnecessary information. The generation of mask $\bm{m}\in\{0,255\}^{H\times W\times C}$ can be depicted as follows: we first apply PT on one image with all 255 pixel value according to $\bm{h}$, and then take logical negation operation on the transformed image, which is illustrated in Fig. \ref{mask}. After that, $\bm{m}$ will conduct pixel-wise production with $\bm{x}$ to complete the masking.

%\vspace{0.4cm}

However, a non-learnable PT process can hardly handle the case of varying homography, which commonly happens in the real world owing to the changing camera poses, illuminations, shooting parameters as well as movements of objects. All the reasons enable $\bm{h}$ between captured photos to slightly alter despite the invariant positions, which inevitably affects the inference performance and brings challenges to our proposed framework. The defects result from non-learnable PT lie as the following: first, the algorithm-based methods like SIFT and ORB only work well with enough matched feature points between image pairs. However, for image pairs of large parallax, it is possible to suffer from situations with insufficient feature matching points, which makes descriptors computation challenging and results in an empty homography matrix. In such cases, a fixed masking strategy leads to a very poor reconstruction outcome. The first row in Fig. \ref{Fig_fail} presents a fail recovery of RWZC with a fail homography estimation, in which an incorrect $\bm{h}$ results in an incorrect mask and thus intolerable reconstruction. Secondly, the non-learnable masks cannot endure the slight inaccuracy of $\bm{h}$, while a generated mask with a few errors unavoidably leads to the mismatched blending. This pixel-level-mismatching (labeled in the red rectangular in the second row of Fig. \ref{Fig_fail}) will thus results in significant performance loss, especially for objective metrics.

\begin{figure}[tbp]
	\centering
	\includegraphics[width=0.48\textwidth]{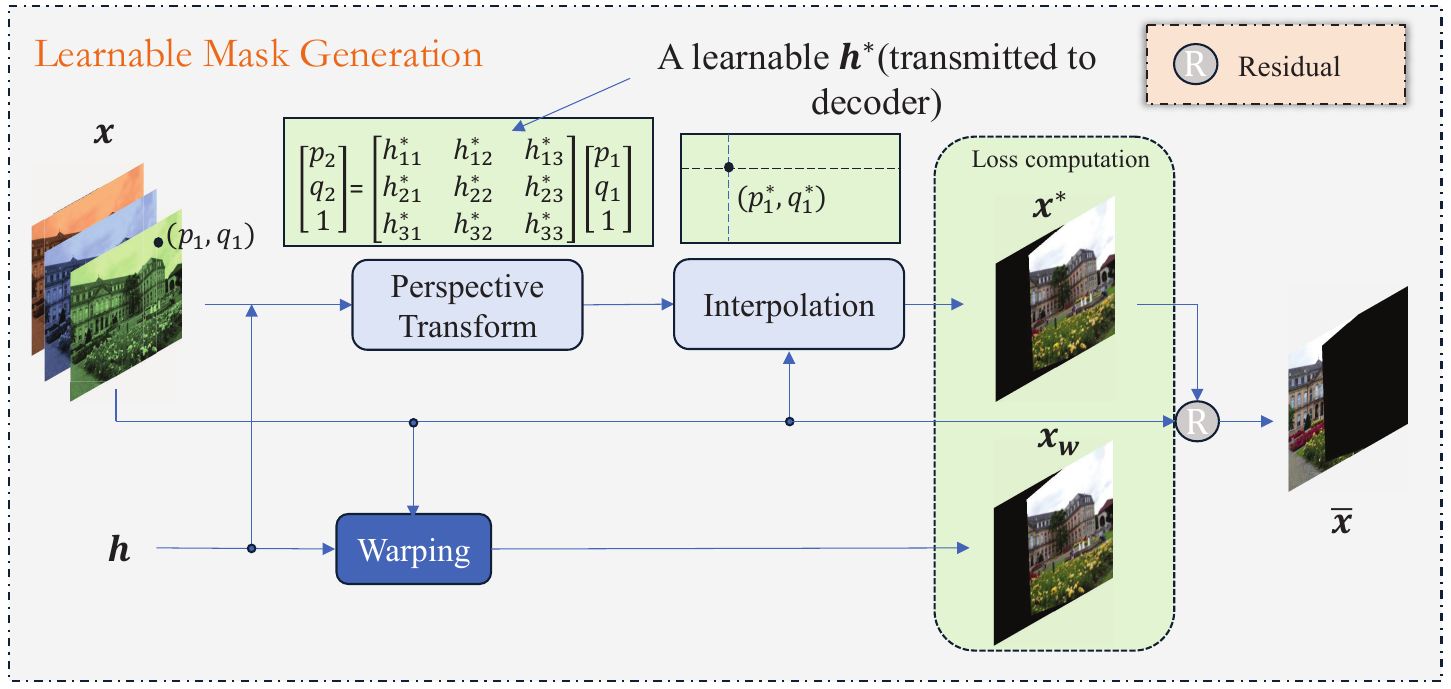}
	\captionsetup{font={small},justification=raggedright}
	\caption{The structure of learning-based $\mathcal{M}_\eta$.}
	\label{deep_homo}
\end{figure}

To tackle the above problem, a learnable mask generation network $\mathcal{M}_\eta$ involving a learning-based homography estimation is considered in our framework. For the deep homography estimation, the pioneer work was proposed by Detone et.al \cite{DeToneMR16}, in which they artificially generated a training set with labels. This supervised network though attained good performance, was restricted for the real world data. An unsupervised one was proposed by Nguyen, et.al \cite{NguyenCSTK18}, which however suffers from performance degradation and high complexity owing to direct linear transforms and spatial decorrelations. 

\begin{figure*}[htbp]
	\centering
	\includegraphics[width=1\textwidth]{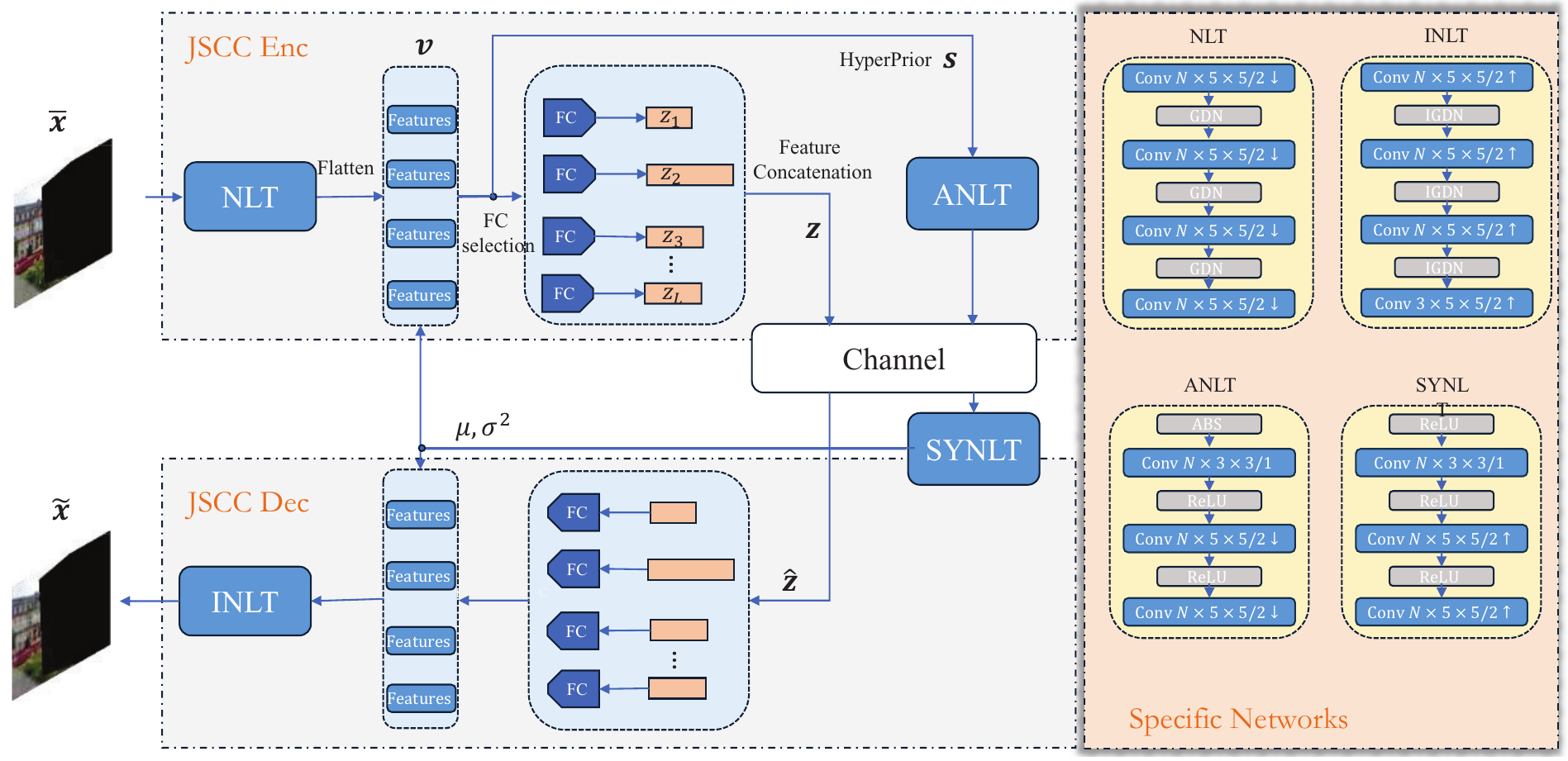}
	\captionsetup{font={small},justification=raggedright}
	\caption{Details for JSCC codec, where $\mathrm{NLT}$ and $\mathrm{INLT}$ represent the non-linear and inverse non-linear transform coding, respectively, while $\mathrm{AnlT}$ and $\mathrm{SynT}$ denote the analysis transform and synthesis transform, respectively.}
	\label{Network}
\end{figure*}

Consequently, we proposed an improved module based on the PTL \cite{DBLP:journals/corr/abs-2201-05706} in this paper. To detail the module, we first clarify the difference between $\bm{h}$ and $\bm{h}^*$ as illustrated in Fig. \ref{deep_homo}, in which the former is the algorithm-calculated homography by $\mathcal{H}$, while the latter is the learnable counterpart used for adaptive masking strategy. Specifically, we first conduct the mask generation in Fig. \ref{mask} and pixel-wised production with $\bm{x}$, thus obtain the target masked image $\bm{x}_w$. Meanwhile the original image $\bm{x}$ is sent into learning-based module conducting PT and interpolations to approach the target. Note that the channel-wised PT operation converts
the homogeneous coordinates to Cartesian coordinates by utilizing $\bm{h}$ as the initial state, and the interpolation step determines each updated value at each new location. Then we train the module by minimizing the distance between the learnable masked image $\bm{x}^*$ and the target masked image $\bm{x}_w$, with the loss computed by
\begin{align}
	\mathcal{L}_M(\eta)=\mathbb{E}_{\bm{x}}\left[d(\bm{x}^*,\bm{x}_w)\right],
\end{align}
where $\bm{x}^*$ and $\bm{x}_w$ are both functions of $\bm{x}$. Finally outputs the residual image $\bar{\bm{x}}=\bm{x}-\bm{x}^*$. At the same time, the learnable $\bm{h}^*$ will replace $\bm{h}$ and be transmitted to the decoder for high-quality reconstruction. In general, this is a special supervised learning that enables the rapid acquisition of SSI while leveraging the generalization capability of neural networks to overcome estimation errors introduced by the algorithm. Besides, the PTL-based masking strategy enjoys the advantages of negligible training overload due to the differentiable linear layer design and multi-viewpoints awareness, namely we can input multiple $\bm{h}$ into the networks for training at one time to realize a better generalization.

\subsection{Entropy model-based transmission}

For the reliable and efficient transmission, we provide a rate-adaptive JSC codec based on hyperprior-assisted non-linear transform. The reason why rate adaption is necessary lies on the masks of irregular sizes, which yield an inefficient transmission if we send images of different masked regions with a same output length. Fortunately, a promising solution is provided by \cite{MinnenBT18} and \cite{DaiWTSQ0022} who uses the entropy model for image transmission with the assistance of latent hyperprior. Therefore, we provide a CNN-based autoencoder for rate-adaptive transmission by incorporating channel coding into the entropy model based framework. The architecture can be found in Fig. \ref{Network}. Specifically, the masked image $\bar{\bm{x}}$ is first converted to features $\bm{v}$ via non-linear transform  composed of convolutional layer and normalization. After that, spatial dependencies is captured by analysis transform on features, and we model the feature latent as scale Gaussian variables via the hyperprior $\bm{s}$, which enables a more precise computation of entropy on each component. Then, the components of features select FC layers of different output dimensions according to the respective computed entropy, and hence realizes an effective rate adaption transmission. At the receiver, the reverse operations are conducted by synthesis transform and inverse non-linear transform modules, and finally the model outputs a reconstructed masked image $\tilde{\bm{x}}$. For the training, the rate-distortion trade-off is utilized as the loss function, which is based on a variational inference, i.e.

\begin{align}
	\mathcal{L}_T(\theta,\xi)=\mathbb{E}_{\bar{\bm{x}}}\left[a\log P_{\bm{v}|\bm{s}}(\bm{v}|\bm{s})+d(\bar{\bm{x}},\tilde{\bm{x}})\right],
\end{align}
The detailed computation can be found in \cite{DaiWTSQ0022}. Note that $\bm{v}$ and $\bm{s}$ are both functions of $\bar{\bm{x}}$, and $a$ denotes the trade-off parameter between rate and distortions. In summary, the integration of adaptive mask generation with an entropy model-based codec achieves both SSI robustness and low-rate transmission. Due to above design, we find that even though RWZC encounters an unfamiliar samples during the inference process, the learning of the distribution with the homography adaption will allow the framework to effectively masking and transmitting $\bm{x}$, which will be shown in the cross-dataset testing later.
\begin{figure*}[tbp]
	\centering
	\includegraphics[width=1\textwidth]{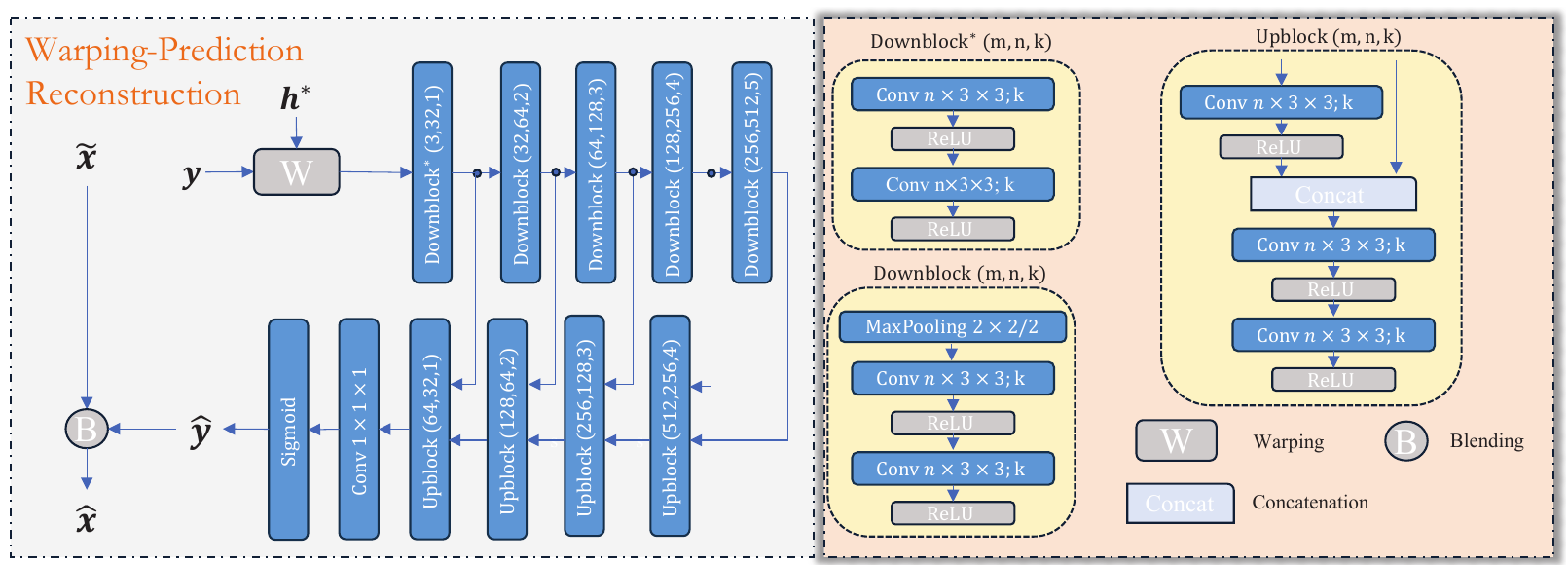}
	\captionsetup{font={small},justification=raggedright}
	\caption{The detailed structure of warping prediction module $\mathcal{W}_\zeta$.}
	\label{decoder_net}
\end{figure*}

\subsection{Warping-predicted reconstruction with SSI}\label{Warp-Pred}

For the high-quality reconstruction, a direct idea is to blend masked image $\tilde{\bm{x}}$ and a warped side image $\bm{y}$. One can find the above modules present a transmitted image portion at the receiver, while the complement portion can also be obtained by the correlated side image $\bm{y}$. However, the pixel-wise blending will inevitably result in two problems. The first problem has been illustrated in the second row in Fig. \ref{Fig_fail} in Sec. \ref{decoupling}, since the varying $\bm{h}$ at the receiver influences the blending performance owing to a distorted perspective transform and an inaccurate SSI.  Second, unlike the pure compression framework, transmission inevitably incurs noise which affects the quality of the received image, and further influences the final combination of $\tilde{\bm{x}}$ and $\bm{y}$. Specifically, due to the resolution differences and noise interference at the masked edges, distortion will occur while ultimately resulting in discontinuities in the reconstructed image, especially for communication environments with extremely low rates and strong interference. These two problems though impacts the perceptual performance slightly, while it is crucial for some objective metrics.
%As illustrated in the second row in Fig. \ref{Fig_fail}, although we can obtain the homography for image pairs of moderate parallax at the training phase, the non-learnable warping at the decoder cannot be cope with the changeable $\bm{h}$ at the testing phase. This pixel-level-mismatching (labeled in the red rectangular in Fig. \ref{Fig_fail}) will inevitably results in significant performance loss, especially for Peak Signal-to-Noise Ratio (PSNR).

\begin{figure}[htbp]
	\centering
	\subfloat[original image]{
		\includegraphics[width=0.15\textwidth]{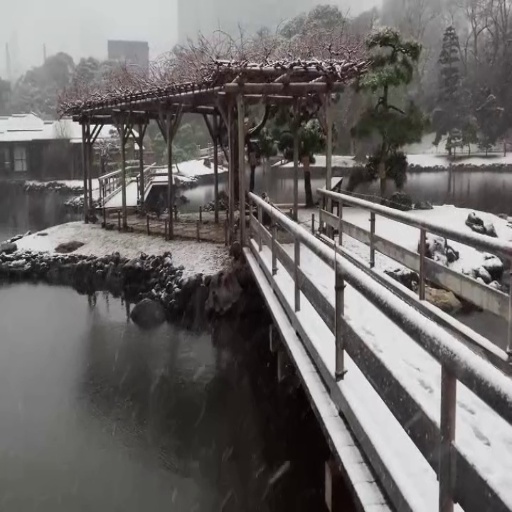}}
	\subfloat[without $\mathcal{W}_\zeta$]{
		\includegraphics[width=0.15\textwidth]{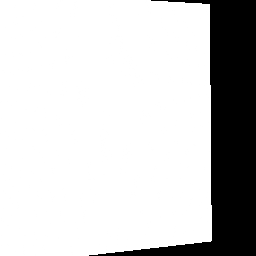}}
	\subfloat[imperfect blending]{
		\includegraphics[width=0.151\textwidth]{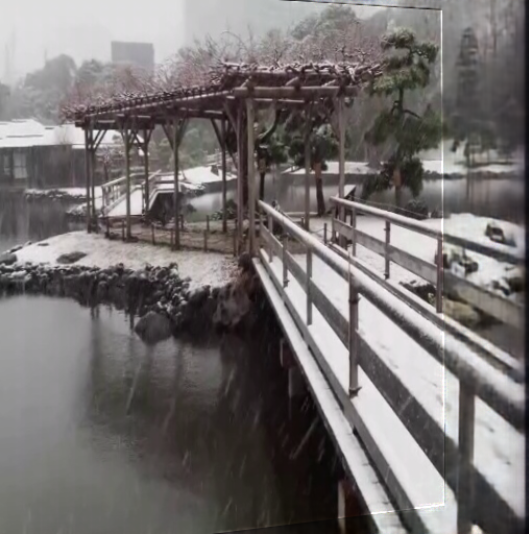}}
	\newline
	\centering
	\subfloat[side image]{
		\includegraphics[width=0.15\textwidth]{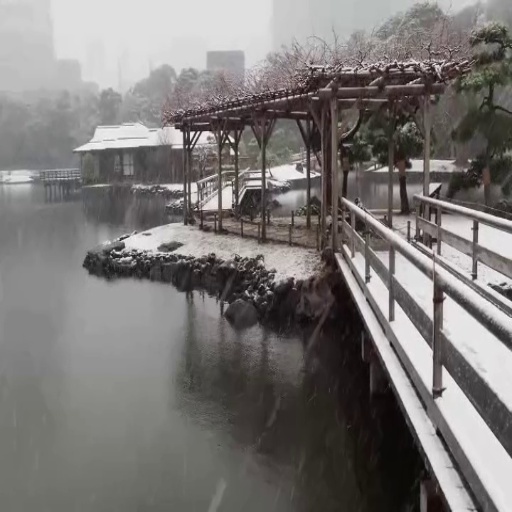}}
	\subfloat[with $\mathcal{W}_\zeta$]{
		\includegraphics[width=0.15\textwidth]{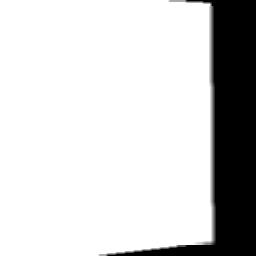}}
	\subfloat[seamless blending]{
		\includegraphics[width=0.151\textwidth]{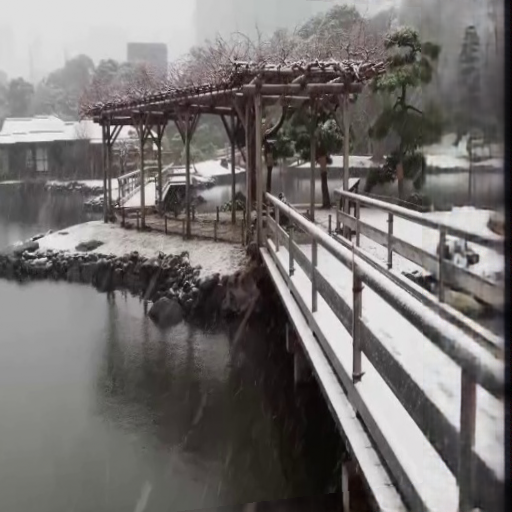}}
	\captionsetup{font={small},justification=raggedright}
	\caption{Comparison of RWZC w/o $\mathcal{W}_\zeta$}\label{Fig_seam}
\end{figure}

%\vspace{0.2cm}
To fix the above problem, we thus propose a learnable reconstruction network to predict the warping operation with the help of SSI. Specifically, we aim to generate a warped side image from $\bm{y}$ to complement the received image portion. To do this, we incorporate an improved module based on UNet \cite{Ronneberger_Fischer_Brox_2015} for channel-wise mask prediction and thus pixel compensation. The detailed structure is shown in Fig. \ref{decoder_net}, where side image $\bm{y}$ is first warped by learnable SSI $\bm{h}^*$, and then fed into $5$ successive 'Downblock' modules composed of Maxpooling and convolutional layers. As a target, this segmentation-oriented module predicts a proper masking for reconstruction through $4$ deconvolutional layers. Here, a novel operation is that the decoder employs a latent variable embedding structure guided by the optimal transform $\bm{h^*}$, thus integrating SSI into the warped $\bm{y}$ and generating features to adapt the mismatch caused by noise and inaccurate estimation. After the sigmoid, this module outputs a complementary image $\hat{\bm{y}}$, which is used to compensate the pixels around the stitching seam affected by resolution difference and noise. The loss function can be formulated as
\begin{align}
	\mathcal{L}_{W}(\zeta) = \mathbb{E}_{\bm{x},\bm{y}}\left[d(\bm{x},\hat{\bm{x}})+b\mathcal{L}_{s}\right],
\end{align}
where $\mathcal{L}_{s}$ denotes the boundary and smoothness loss \cite{NieLLLZ23} for seamless blending, while $b$ is the parameter to trade-off these two terms. Fig. \ref{Fig_seam} shows the effect of the learning-based $\mathcal{W}_\eta$, in which the first row completes the deirect blending with a perspective transformed side image according to $\bm{h}$. One can easily find the obvious pixel deviation at the stitching position caused by resolution difference of two portions. In general, this problem can be solved by the predicted compensation and blur operation around the seam via our network.

%\restoregeometry
\subsection{Loss function and training algorithm}
In this work, our proposed framework has three main learning-based neural networks, namely adaptive masking $\mathcal{M}_{\eta}$, JSC codec $F_{\theta}$ and $F_{\xi}$, and warping prediction reconstruction $\mathcal{W}_{\zeta}$. For the training process, we jointly train the three networks for an optimal performance, in which the combination of PTL-based mask and variational autoencoder (VAE)-based JSC codec ensures the performance against varying homography and channel interference, while prediction-based warping completes a pixel-level refinement. Moreover, we remark that the joint training though suffers from a higher complexity with its separated counterpart, it always enjoys a better performance. In specific, 
%In stage 2, we apply warping prediction to improve the final performance and realize an efficient reconstruction framework. The stage-wise training benefits from a lower temporal complexity and higher generalization in comparison with a joint training of all three networks, at the cost of possible better performance. 
the loss functions can be formulated as follows.
\begin{align}
	\mathcal{L}(\theta,\xi,\eta,\zeta)=\mathcal{L}_M(\eta)+\lambda_1\mathcal{L}_T(\theta,\xi)+\lambda_2\mathcal{L}_W(\zeta),
\end{align}
where $\lambda_1$ and $\lambda_2$ are the hyperparameters to trade-off homography-robustness, decoupling performance, and the warping predictions. The distortion measure $d(\cdot,\cdot)$ can be selected from peak signal-to-noise ratio (PSNR) and multi-scale structural similarity (MS-SSIM). The detailed training algorithm can be found in Algorithm \ref{Alg1}.
%\IncMargin{1em}
\begin{algorithm}
	\caption{Training for RWZC}
	\SetKwData{Left}{left}\SetKwData{This}{this}\SetKwData{Up}{up}
	\SetKwFunction{Union}{Union}\SetKwFunction{FindCompress}{FindCompress}
	\SetKwInOut{Input}{Input}\SetKwInOut{Output}{Output}
	\SetKwRepeat{Repeat}{repeat}{until}
	\Input{Original image $\bm{x}$, side image $\bm{y}$, channel SNR and maximal epochs $i_{\mathrm{max}}$}
	\BlankLine
	\label{Alg1}
	\Repeat{$i_{\mathrm{max}}$ is reached}{
		Shuffling and randomly flipping $D_{\mathrm{train}}$, set $i=0,L=0$
		\BlankLine
		\For{$(\bm{x},\bm{y})\in D_{\mathrm{train}}$}{
			\BlankLine
			$\bm{h}=\mathcal{H}(\bm{x},\bm{y})$\tcc*[h]{\footnotesize Homography estimation}
			\BlankLine
			$(\bar{\bm{x}},\bm{h}^*)=\mathcal{M}_\eta(\bm{x},\bm{h})$;\tcc*[h]{\footnotesize Adaptive masking}
			\BlankLine
			$\bm{z} = F_{\theta}(\bm{x})$;\tcc*[h]{\footnotesize JSC encoder}
			\BlankLine
			$\hat{\bm{z}} = \bm{z}+\bm{n}$;\tcc*[h]{\footnotesize Randomly noise via SNR}
			\BlankLine
			$\tilde{\bm{x}}=F^{-1}_{\xi}(\hat{\bm{z}})$
			\BlankLine
			$\hat{\bm{x}}=\mathcal{W}_{\zeta}(\tilde{\bm{x}},\bm{y},\bm{h}^*)$\tcc*[h]{\footnotesize Reconstruction}
			\BlankLine
			$L = L+\mathcal{L}(\theta,\xi,\eta,\zeta)$;\tcc*[h]{\footnotesize Loss accumulation}
			\BlankLine
			Update $\theta,\xi,\eta,\zeta$ via backpropagating $L$
		}
	}	
\end{algorithm}
%\DecMargin{1em} 
\section{Experiments}\label{Sec5}
In this section, performance of our proposed framework will be verified. We select two state-of-the-art competitors, which  consider efficient compression and channel-robust transmission at a WZ coding scenario, respectively. In conclusion, for the reconstruction task of $\bm{x}$, our framework outperforms the baselines on perceptual metrics significantly, while maintaining comparable performance on objective metrics, especially for extremely low rate.
%For the improved framework HAJSCC-SSI, ablations are provided to show the superiority to its original version. 
Moreover, comparison of complexity is presented to demonstrate the advantage of the light-weight codec design in our work.

\subsection{Experimental Setup}
We present the experimental settings in this subsection, including the used datasets, parameter details, the competitors and performance metrics.
\subsubsection{Datasets}
Two datasets are considered in this work. The first is the well-known KITTI Stereo Dataset \cite{GeigerLU12,Menze15} in the fields of computer vision and machine learning, consisting of 8000 stereo image pairs, in which we use 4000 for training, 3250 and 750 for testing and validation. Note that KITTI Stereo is a small and \textbf{stationary parallax} dataset, thus it probably enables a virtually unchanged homography. This hence motivates us to consider another stereo image dataset named UDIS-D \cite{NieLLLZ21} of \textbf{varying parallax}, which is proposed by Lang et al. for unsupervised image stitching. This dataset is composed of image pairs from frames in videos and self-taken photos, resulting in an irregular $\bm{h}$, which is more suitable for the performance validation. For UDIS-D, we adopt 8334 pairs for training, 1107 and 1000 pairs for testing and validation respectively. 

\subsubsection{Evaluation Metrics}
To evaluate the performance of the decoder-only side information empowered framework, we adopt the widely-used metric \textbf{PSNR}, which is used to evaluate the quality of reconstructed images compared to the original image. Except for the point-wised object metric, we also used \textbf{MS-SSIM} and learned perceptual image patch similarity (\textbf{LPIPS}) for evaluation, which are both used as the proxy for the human perception. Among them, MS-SSIM is calculated with luminance, contrast, and structure between two images, while LPIPS computes the distance in the feature space based on a pretrained network.

\subsubsection{Model and Training Details}
We model the complex AWGN channel with noise sampled from $\mathcal{CN}(\bm{0},\sigma_n^2\bm{I}_k)$, normalize the transmitting power and control $\sigma_n^2$ for SNR computation. 
In RWZC framework, we adopt algorithms SURF and RANSAC 5.0 for feature matching and defects exclusion in $\mathcal{H}$, respectively. For mask generating strategy $\mathcal{M}_\eta$, we choose bicubic interpolation for PIL and the list size of trainable homography $\bm{h}^*$ equals to 1. For JSCC codec $F_{\theta}$ and $F^{-1}_{\xi}$, input/output/kernal details are labeled in Fig. \ref{Network}, and we construct a set of 16 different FC layers, whose output size ranges from 4 to 128 to realize the transmission rate adaption. For final stitching in $\mathcal{W}_\zeta$, Poisson blending is used for seamless reconstruction. Moreover, we resize the images in 370$\times$740 and 256$\times$256 for KITTI Stereo and UDIS-D datasets, respectively. Besides, for the training, $i_{\mathrm{max}}$ equals to 400, $\lambda_1=0.7$ and $\lambda_2=0.1$ while the learning rate is set as $1e^{-4}$. The framework is optimized with Adam \cite{KingmaB14} and trained with the distortion measure of MS-SSIM.

\subsubsection{Baselines}
Besides RWZC, we also select a transmission efficient framework and a compression based framework for performance comparison, namely Deep Joint Source-Channel coding-Wyner Ziv (DeepJSCC-WZ, by Yilmaz. et al.\cite{YilmazKG23_2}) and Neural distributed image compression (NDIC, by Mital et al. \cite{MitalOGG22}) combined with capacity achieving channel coding. Details can be found in the following.
\\\textbf{DeepJSCC-WZ} is a state-of-the-art low-latency image transmission framework with decoder-only side information. This data-driven approach involves attention feature module into the multi-stages design and SNR-adaptive mechanism, thus offers graceful performance on image reconstruction and efficient transmission. Moreover, DeepJSCC-WZ provides the rate choices of $\frac{1}{16}$ and $\frac{1}{32}$, yielding two points in the following plots which is against CBR.
\\\textbf{Mital+capacity} denotes the scheme of NDIC combined with capacity achieving code. It is a separation-based framework, in which the compression is based on the entropy model from \cite{BalleCMSJAHT21} with a common feature extraction at decoder, and the transmission is assumed error-free at the maximun achievable rate, computed by $\frac{k}{n}$ times the channel capacity\cite{DBLP:journals/tccn/BourtsoulatzeKG19}. Specifically, the method of Mital et al. utilizes learned image compression with hyperprior as the backbone and loss function characterizing the trade-off between rate and distortion. It should be noticed that, in Mital+capacity, the CBR cannot be less than $0.027$ since too strict compression may lead to non-convergence during the training process.
\begin{figure*}[htbp]
	\centering
	\begin{minipage}[t]{0.312\linewidth}
		\centering
		\includegraphics[width=1\textwidth]{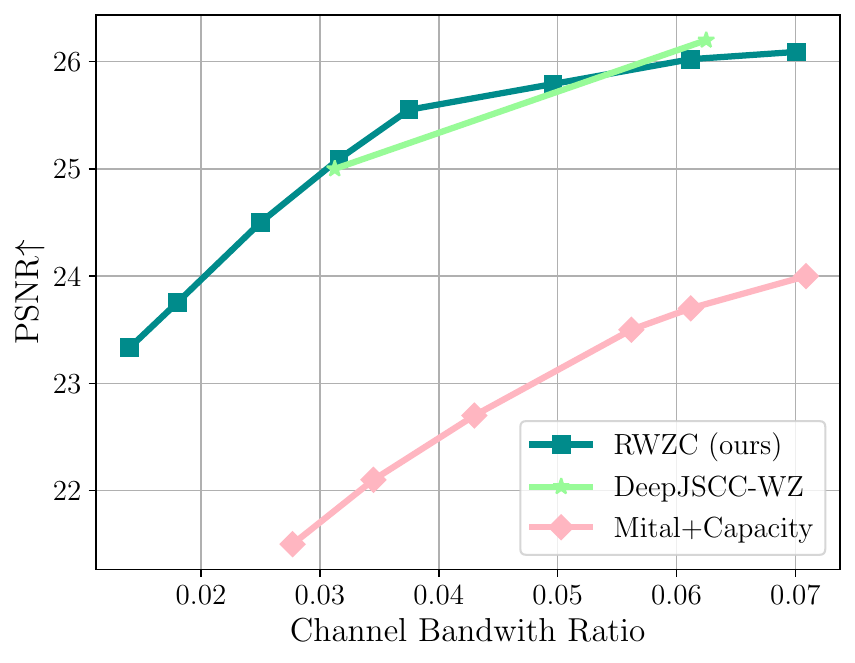}
		%		\vspace*{-35ex} 
		%		\begin{center}
			%			KITTI Stereo/SNR=3
			%		\end{center}
		%		\vspace*{34ex}
	\end{minipage}
	\begin{minipage}[t]{0.32\linewidth}
		\centering
		\includegraphics[width=1\textwidth]{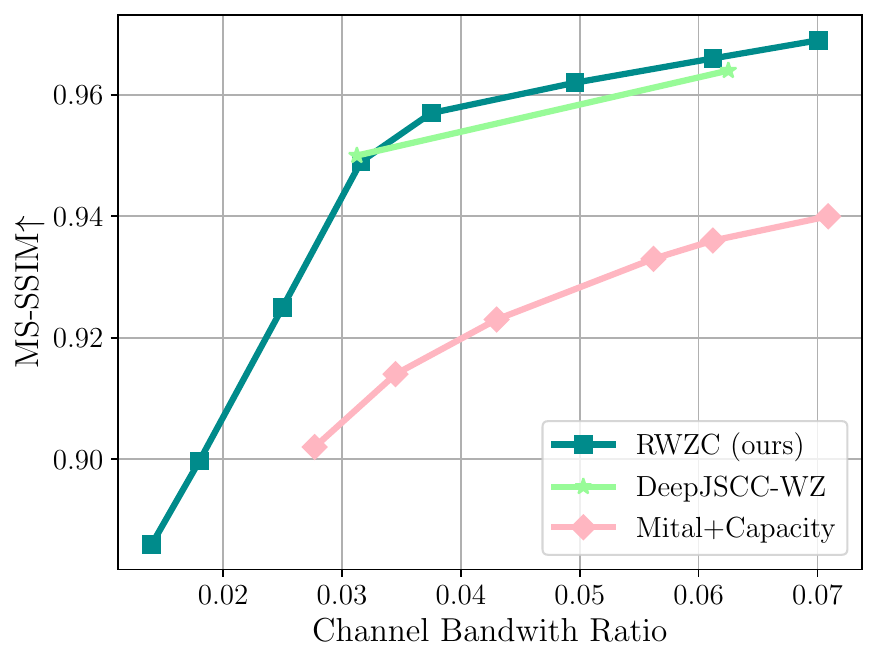}
		%		\vspace*{-35ex} 
		%		\begin{center}
			%			KITTI Stereo/SNR=3
			%		\end{center}
		%		\vspace*{16ex}
	\end{minipage}
	\begin{minipage}[t]{0.32\linewidth}
		\centering
		\includegraphics[width=1\textwidth]{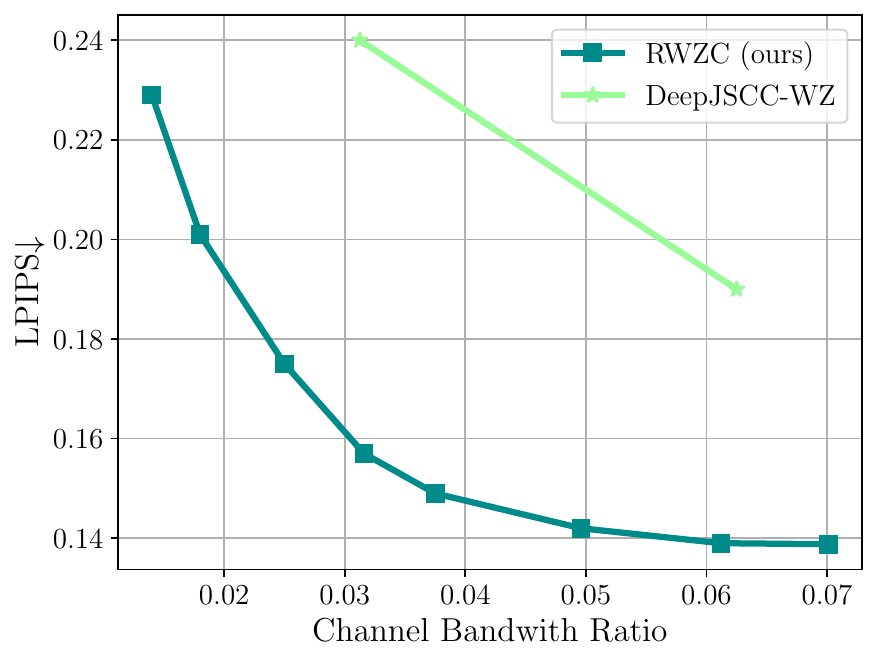}
		%		\vspace*{-35ex} 
		%		\begin{center}
			%			KITTI Stereo/SNR=3
			%		\end{center}
		%		\vspace*{16ex}
	\end{minipage}
	\centering
	%\vspace*{-6ex}
	\captionsetup{font={small},justification=raggedright}
	\caption{Comparison of proposed method with baselines on the KITTI stereo dataset by fixing SNR=3.}\label{KITTI_CBR}
\end{figure*}
\begin{figure*}[htbp]
	\centering
	\begin{minipage}[t]{0.31\linewidth}
		\centering
		\includegraphics[width=1\textwidth]{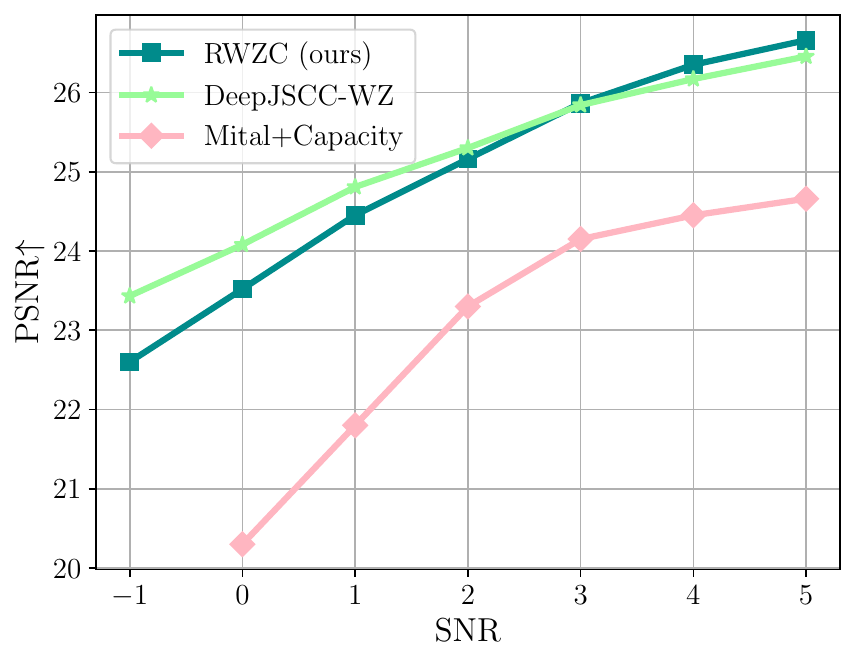}
		%		\vspace*{-35ex} 
		%		\begin{center}
			%			KITTI Stereo/CBR$\approx$0.031
			%		\end{center}
		%		\vspace*{34ex}
	\end{minipage}
	\begin{minipage}[t]{0.32\linewidth}
		\centering
		\includegraphics[width=1\textwidth]{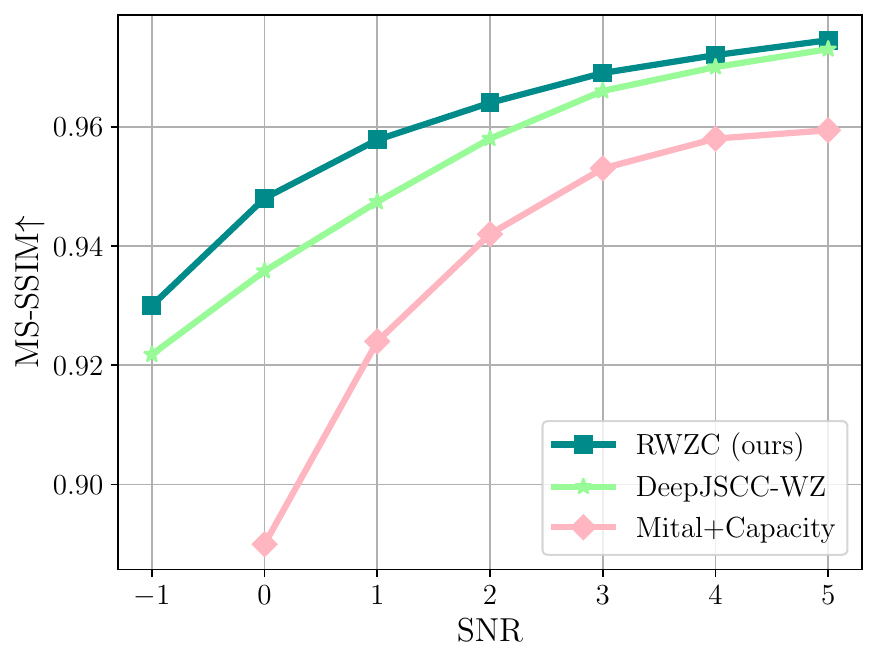}
		%		\vspace*{-35ex} 
		%		\begin{center}
			%			KITTI Stereo/CBR$\approx$0.031
			%		\end{center}
		%		\vspace*{16ex}
	\end{minipage}
	\begin{minipage}[t]{0.32\linewidth}
		\centering
		\includegraphics[width=1\textwidth]{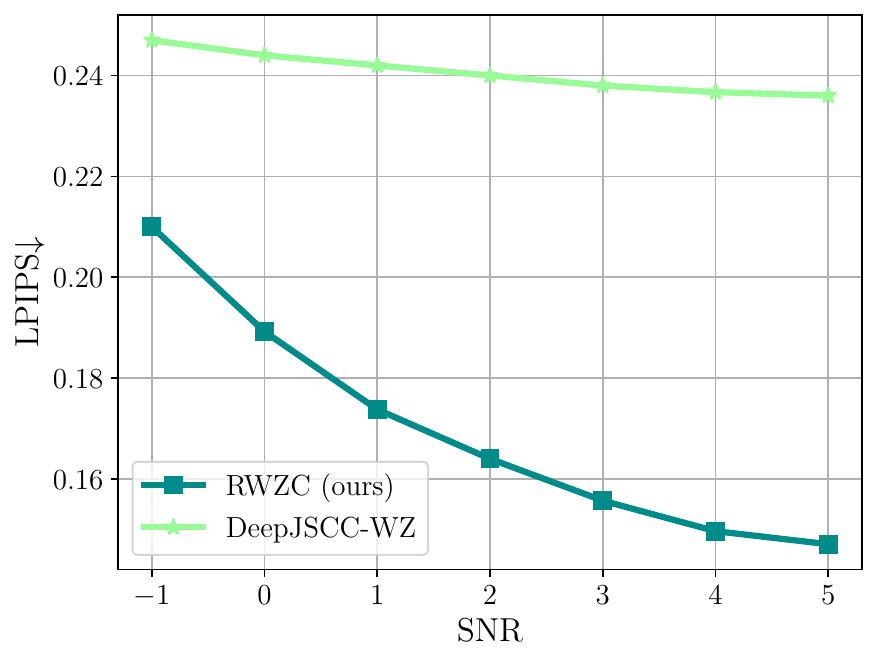}
		%		\vspace*{-35ex} 
		%		\begin{center}
			%			KITTI Stereo/CBR$\approx$0.031
			%		\end{center}
	\end{minipage}
	%\vspace*{-6ex} 
	\captionsetup{font={small},justification=raggedright}
	\caption{Comparison of proposed method with baselines on the KITTI stereo dataset by fixing CBR$\approx0.031$.}\label{KITTI_SNR}
\end{figure*}
\begin{figure*}[htbp]
	\centering
	\begin{minipage}[t]{0.13\linewidth}
		\centering
		\vspace*{7ex}
		\includegraphics[width=1\textwidth]{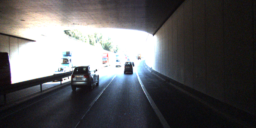}
		\vspace*{-17ex} 
		\begin{center}
			\textbf{Image $\bm{x}$}
		\end{center}
		\vspace*{9ex}
		\begin{center}
			\small CBR\\
			PSNR/SSIM/LPIPS
		\end{center}
		\vspace*{1ex}
	\end{minipage}
	\begin{minipage}[t]{0.13\linewidth}
		\centering
		\vspace*{7ex}
		\includegraphics[width=1\textwidth]{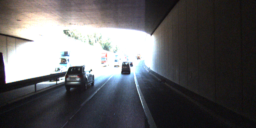}
		\vspace*{-17ex} 
		\begin{center}
			\textbf{Side image $\bm{y}$}
		\end{center}
	\end{minipage}
	\quad
	\begin{minipage}[t]{0.13\linewidth}
		\centering
		\vspace*{7ex}
		\includegraphics[width=1\textwidth]{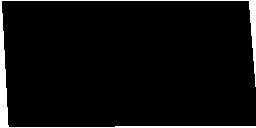}
		\vspace*{-17ex} 
		\begin{center}
			\textbf{Masking (SSI)}
		\end{center}
	\end{minipage}
	\quad
	\begin{minipage}[t]{0.13\linewidth}
		\centering
		\vspace*{7ex}
		\includegraphics[width=1\textwidth]{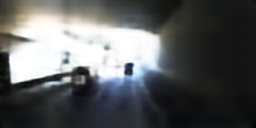}
		\vspace*{-17ex} 
		\begin{center}
			\textbf{Mital+Cap}
		\end{center}
		\vspace*{9ex}
		\begin{center}
			\textbf{\small$0.0683$\protect\\
				$27.43/0.85$}
		\end{center}
	\end{minipage}
	\quad
	\begin{minipage}[t]{0.13\linewidth}
		\centering
		\vspace*{7ex}
		\includegraphics[width=1\textwidth]{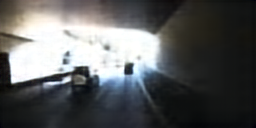}
		\vspace*{-17ex} 
		\begin{center}
			\textbf{DeepJSCC-WZ}
		\end{center}
		\vspace*{9ex}
		\begin{center}
			\textbf{\small$0.0625$\protect\\
				$\textbf{29.33}/\textbf{0.938}/0.188$}
		\end{center}
	\end{minipage}
	\quad
	\hspace{0.3cm}
	\begin{minipage}[t]{0.13\linewidth}
		\centering
		\vspace*{7ex}
		\includegraphics[width=1\textwidth]{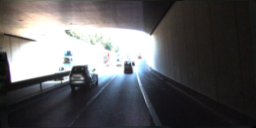}
		\vspace*{-17ex} 
		\begin{center}
			\textbf{RWZC}
		\end{center}
		\vspace*{9ex}
		\begin{center}
			\textbf{\small$0.0601$\\
				$28.92/0.927/\textbf{0.099}$}
		\end{center}
	\end{minipage}
	\vspace*{1ex}
	\begin{minipage}[t]{0.13\linewidth}
		\centering
		\includegraphics[width=1\textwidth]{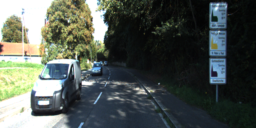}
	\end{minipage}
	\begin{minipage}[t]{0.13\linewidth}
		\centering
		\includegraphics[width=1\textwidth]{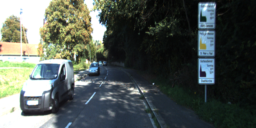}
	\end{minipage}
	\quad
	\begin{minipage}[t]{0.13\linewidth}
		\centering
		\includegraphics[width=1\textwidth]{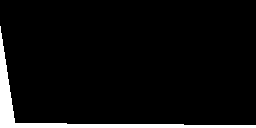}
	\end{minipage}
	\quad
	\begin{minipage}[t]{0.13\linewidth}
		\centering
		\includegraphics[width=1\textwidth]{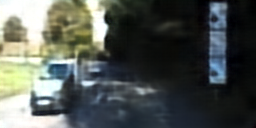}
		%	\vspace*{16ex}
		\begin{center}
			\textbf{\small$0.0709$\\
				$23.37/0.795$}
		\end{center}
	\end{minipage}
	\quad
	\begin{minipage}[t]{0.13\linewidth}
		\centering
		\includegraphics[width=1\textwidth]{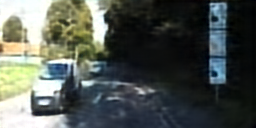}
		\begin{center}
			\textbf{\small$0.0625$\\
				$25.25/\textbf{0.922}/0.29$}
		\end{center}
	\end{minipage}
	\quad
	\hspace{0.3cm}
	\begin{minipage}[t]{0.13\linewidth}
		\centering
		\includegraphics[width=1\textwidth]{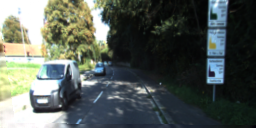}
		\begin{center}
			\textbf{\small$0.0612$\\
				$\textbf{26.84}/0.922/\textbf{0.109}$}
		\end{center}
	\end{minipage}
	\captionsetup{font={small},justification=raggedright}
	\caption{Visualization of the comparison on KITTI Stereo dataset; the optimal performance is labeled with bold.}\label{KITTI_VISUAL}
\end{figure*}
\subsection{Evaluation on Stationary Parallax Dataset}

We first evaluate our proposed framework on KITTI Stereo dataset. The image pairs in this dataset are taken by in-car cameras with fixed positions, which appear with almost fixed homography relation between $\bm{x}$ and $\bm{y}$. Before the comparison, we highlight that all schemes are trained with MS-SSIM measure and are evaluated across different quality metrics, e.g. PSNR, MS-SSIM and LPIPS, and compared within an interval of relative low CBR (0.01-0.07 channel use/symbol).

Fig. \ref{KITTI_CBR} presents the reconstruction performance of all frameworks over different bandwidth usages on the dataset with stationary parallax. In general, our method performs as good as DeepJSCC-WZ and outperforms Mital+capacity, in terms of PSNR and MS-SSIM metrics. Specifically, the PSNR of RWZC can reach 25.5 dB when CBR is around 0.03 (at 3 dB channel SNR), while DeepJSCC-WZ has a slightly weaker performance of 25 dB and Mital+capacity only obtains 22.1 dB. The case turns a bit different when CBR climbs to 0.06, while DeepJSCC-WZ overtakes our scheme around 0.02 dB. Moreover, the MS-SSIM of RWZC achieves 0.956 while DeepJSCC-WZ has 0.952 and Mital+capacity has only 0.918, when CBR approaches 0.03. Besides, our method shows an overwhelming trends compared with DeepJSCC-WZ in LPIPS metric, which appears an average 0.08 improvement as shown in the third plot. 

Fig. \ref{KITTI_SNR} plots the reconstruct performance of all competitors against different channel environments with a fixed CBR. One can find that the overall trends remain almost similar as depicted above. Something different is in the first plot in Fig. \ref{KITTI_SNR}, where DeepJSCC-WZ has a better PSNR performance than RWZC at a worse channel environment (SNR$\leq3$) due to its generalization over different SNR ranges. For the left comparison, on both MS-SSIM and LPIPS, RWZC overwhelms its competitors.
%\vspace*{6ex} 

\begin{figure*}[tbp]
	\centering
	\begin{minipage}[t]{0.31\linewidth}
		\centering
		\includegraphics[width=1\textwidth]{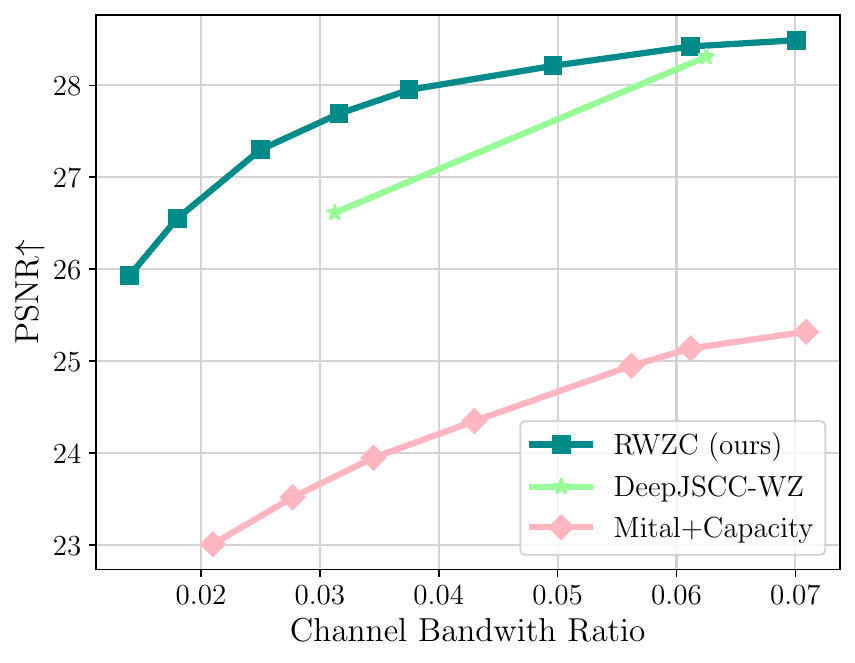}
		%		\vspace*{-37ex} 
		%		\begin{center}
			%			UDIS-D/SNR=3
			%		\end{center}
		%		\vspace*{34ex}
	\end{minipage}
	\begin{minipage}[t]{0.32\linewidth}
		\centering
		\includegraphics[width=1\textwidth]{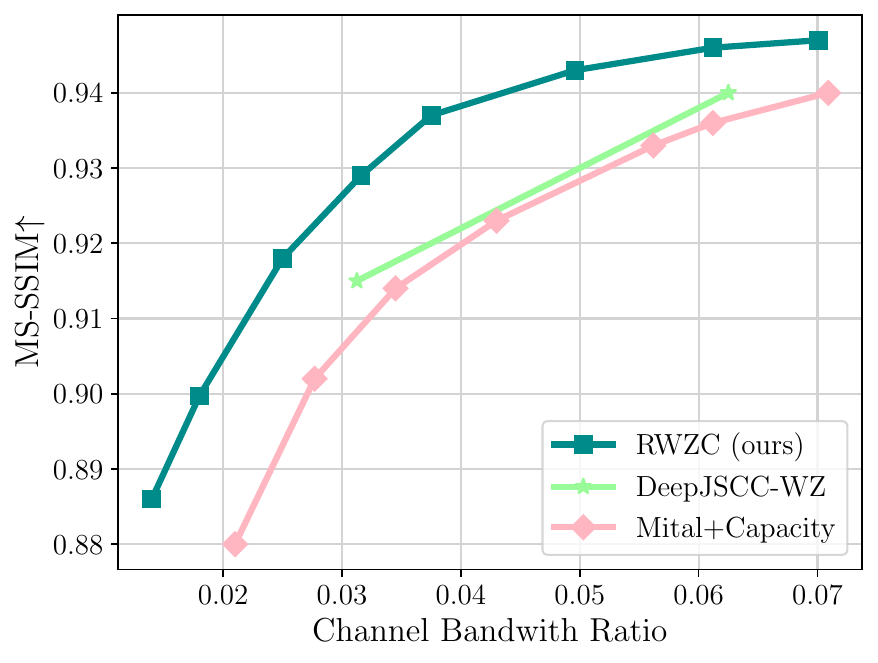}
		%		\vspace*{-37ex} 
		%		\begin{center}
			%			UDIS-D/SNR=3
			%		\end{center}
		
	\end{minipage}
	\begin{minipage}[t]{0.32\linewidth}
		\centering
		\includegraphics[width=1\textwidth]{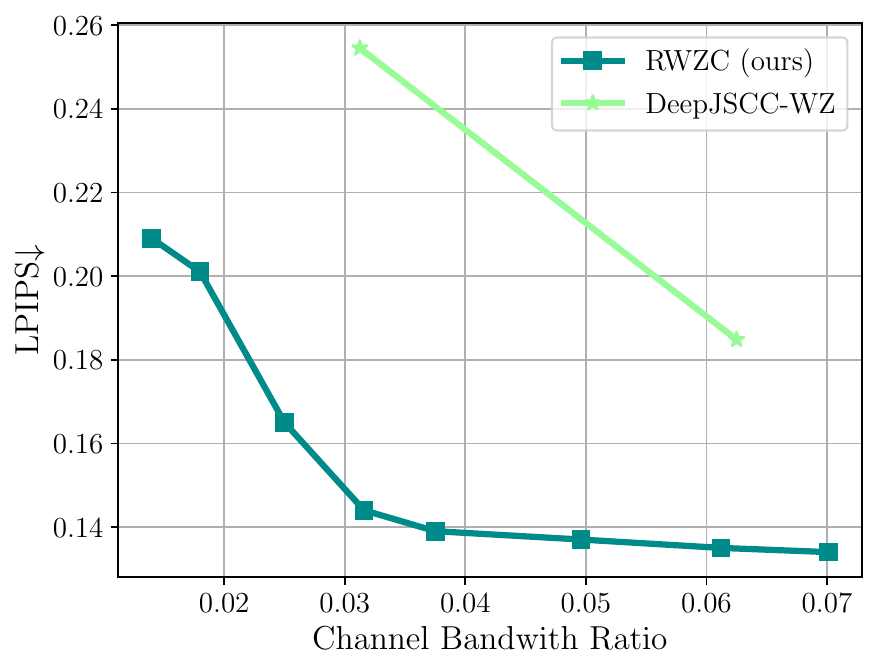}
		%		\vspace*{-37ex} 
		%		\begin{center}
			%			UDIS-D/SNR=3
			%		\end{center}
	\end{minipage}
	\captionsetup{font={small},justification=raggedright}
	\caption{Comparison of proposed method with the baselines on the UDIS-D dataset by fixing SNR=3.}\label{UDIS_CBR}
\end{figure*}
\begin{figure*}[htbp]
	\centering
	\begin{minipage}[t]{0.31\linewidth}
		\centering
		\includegraphics[width=1\textwidth]{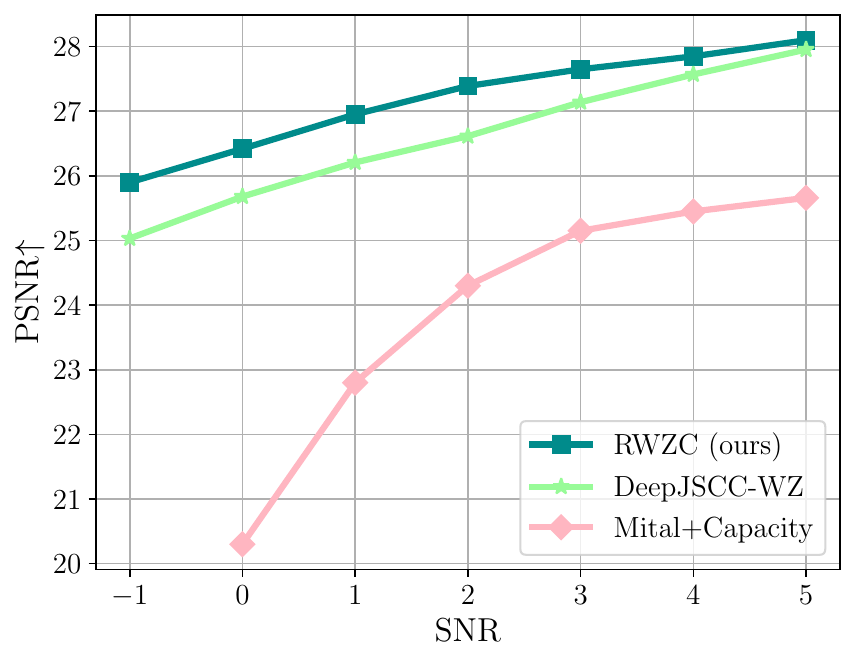}
		%		\vspace*{-37ex} 
		%		\begin{center}
			%			UDIS-D/CBR$\approx$0.031
			%		\end{center}
		%		\vspace*{16ex}
	\end{minipage}
	\begin{minipage}[t]{0.32\linewidth}
		\centering
		\includegraphics[width=1\textwidth]{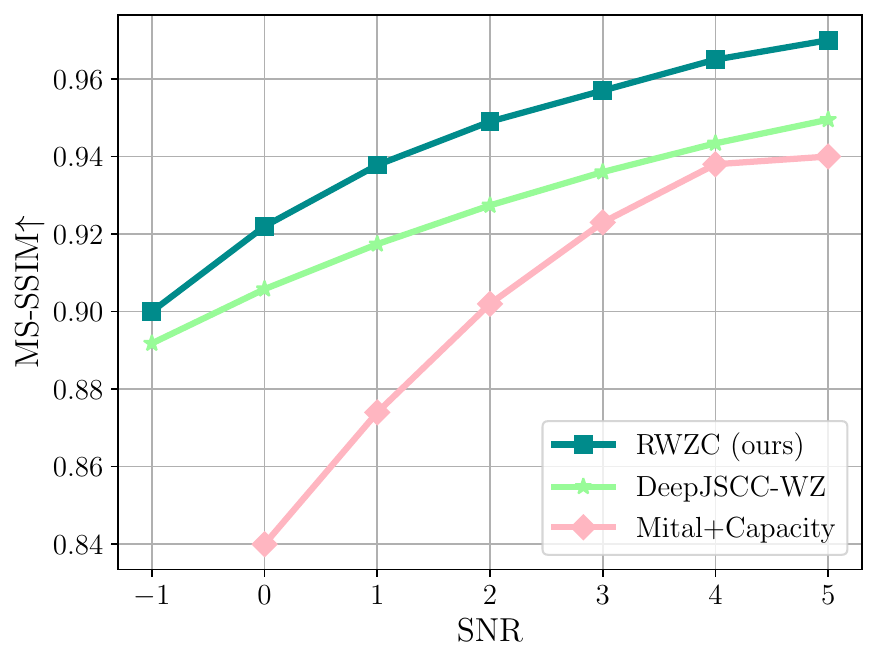}
		%		\vspace*{-37ex} 
		%		\begin{center}
			%			UDIS-D/CBR$\approx$0.031
			%		\end{center}
		%		\vspace*{16ex}
	\end{minipage}
	\begin{minipage}[t]{0.325\linewidth}
		\centering
		\includegraphics[width=1\textwidth]{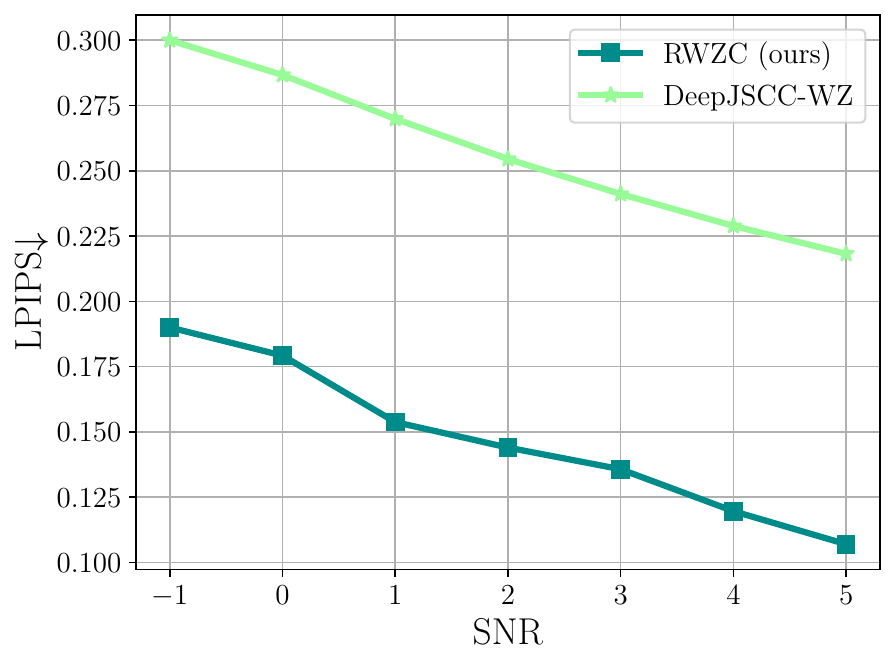}
		%		\vspace*{-37ex} 
		%		\begin{center}
			%			UDIS-D/CBR$\approx$0.031
			%		\end{center}
		%		\vspace*{16ex}
	\end{minipage}
	\captionsetup{font={small},justification=raggedright}
	\caption{Comparison of proposed method with the baselines on the UDIS-D dataset by fixing CBR$\approx0.031$.}\label{UDIS_SNR}
\end{figure*}
The performance degradation of baselines mainly comes for the following reason. In such a low-rate transmission, the feature matching between $\bm{x}$ and $\bm{y}$ is inefficient and difficult due to the small receptive field of the CNN architectures. The different performance gaps are attributed to that DeepJSCC-WZ adopts an attention feature module to extend the receptive field, while Mital+capacity even suffers from sup-optimality owing to the separated source-channel design. Meanwhile, the inefficient matching will not happen in our framework, since we divide the correlated information at the transmitter according to affine transform, which incurs merely interactions on the pixel level at the decoder. Furthermore, the affine transform retains the geometric information of the original image to the maximum extent despite the slightly pixel shifting, which is highly friendly to the human perception. This hence resulting in the superiority of RWZC in LPIPS.

In Fig. \ref{KITTI_VISUAL}, visual results are presented to show the specific performance. The stationary homography is illustrated in the third column (marked as Masking(SSI)), where the $\bm{h}$-generated mask remains almost invariant for most image pairs. The visualization also validates the comparable performance of RWZC with DeepJSCC-WZ. One can conclude that for the KITTI stereo dataset, our proposed method does not obtain a significant advantage when compared to existing methods, (even negative gain on PSNR/MS-SSIM compared with DeepJSCC-WZ) except for the perceptual metric, no matter against CBR or SNR. The reason lies on that the adequate information carried by the side information in the stationary parallax scenario, which offers effective improvement for data-driven-based feature extraction. However, we want to clarify that it is not the same case when the parallax is not stationary, which will be shown in the next section.

\subsection{Evaluation on Parallax-Varying Dataset}

\begin{figure*}[htbp]
	\centering
	\vspace*{5ex}
	\begin{minipage}[t]{0.13\linewidth}
		\centering
		%\vspace*{7ex}
		
		\includegraphics[width=1\textwidth]{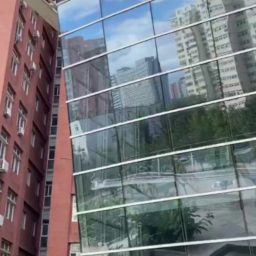}
		\vspace*{-23ex} 
		\begin{center}
			\textbf{Image $\bm{x}$}
		\end{center}
		\vspace*{16ex}
		\begin{center}
			\small CBR\\
			PSNR/SSIM/LPIPS
		\end{center}
	\end{minipage}
	\begin{minipage}[t]{0.13\linewidth}
		\centering
		\includegraphics[width=1\textwidth]{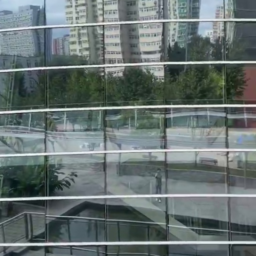}\label{AWGN_Rec_Para}
		\vspace*{-23ex} 
		\begin{center}
			\textbf{Side image $\bm{y}$}
		\end{center}
	\end{minipage}
	\quad
	\begin{minipage}[t]{0.13\linewidth}
		\centering
		\includegraphics[width=1\textwidth]{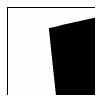}
		\vspace*{-23ex} 
		\begin{center}
			\textbf{Masking (SSI)}
		\end{center}
		\vspace*{16ex}
	\end{minipage}
	\quad
	\begin{minipage}[t]{0.13\linewidth}
		\centering
		\includegraphics[width=1\textwidth]{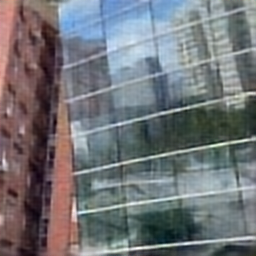}\label{AWGN_Rec_DJ}
		\vspace*{-23ex} 
		\begin{center}
			\textbf{Mital+Cap}
		\end{center}
		\vspace*{16ex}
		\begin{center}
			\textbf{\small$0.0722$\protect\\
				$25.16/0.919$}
		\end{center}
	\end{minipage}
	\quad
	\begin{minipage}[t]{0.13\linewidth}
		\centering
		\includegraphics[width=1\textwidth]{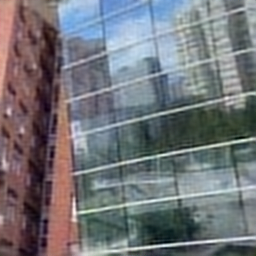}
		\vspace*{-23ex} 
		\begin{center}
			\textbf{DeepJSCC-WZ}
		\end{center}
		\vspace*{16ex}
		\begin{center}
			\textbf{\small$0.0625$\protect\\
				$26.94/0.929/0.26$}
		\end{center}
	\end{minipage}
	\quad
	\hspace{0.3cm}
	\begin{minipage}[t]{0.13\linewidth}
		\centering
		\includegraphics[width=1\textwidth]{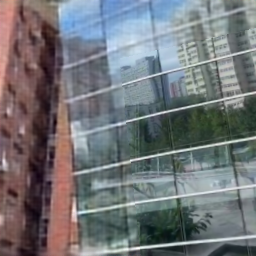}
		\vspace*{-23ex} 
		\begin{center}
			\textbf{RWZC}
		\end{center}
		\vspace*{16ex}
		\begin{center}
			\textbf{\small$0.0619$\\
				$\textbf{28.37}/\textbf{0.942}/\textbf{0.179}$}
		\end{center}
	\end{minipage}

	\vspace*{2ex}
	\begin{minipage}[t]{0.13\linewidth}
		\centering
		\includegraphics[width=1\textwidth]{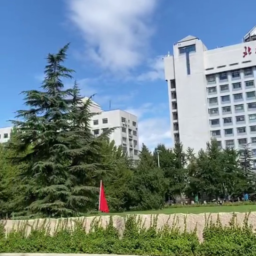}
	\end{minipage}
	\begin{minipage}[t]{0.13\linewidth}
		\centering
		\includegraphics[width=1\textwidth]{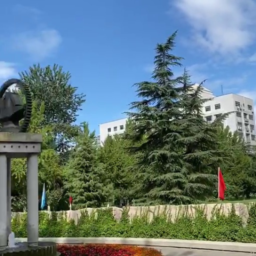}
	\end{minipage}
	\quad
	\begin{minipage}[t]{0.13\linewidth}
		\centering
		\includegraphics[width=1\textwidth]{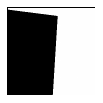}
		\vspace*{-23ex} 
		\vspace*{16ex}
	\end{minipage}
	\quad
	\begin{minipage}[t]{0.13\linewidth}
		\centering
		\includegraphics[width=1\textwidth]{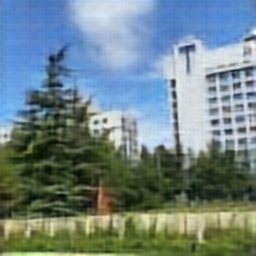}
		%	\vspace*{16ex}
		\begin{center}
			\textbf{\small$0.0709$\\
				$24.37/0.92$}
		\end{center}
	\end{minipage}
	\quad
	\begin{minipage}[t]{0.13\linewidth}
		\centering
		\includegraphics[width=1\textwidth]{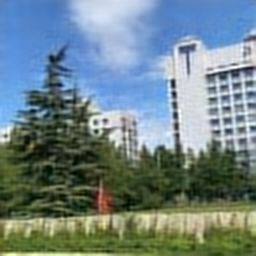}
		\begin{center}
			\textbf{\small$0.0625$\\
				$26.32/0.925/0.24$}
		\end{center}
	\end{minipage}
	\quad
	%\begin{minipage}[t]{0.13\linewidth}
	%	\centering
	%	\includegraphics[width=1\textwidth]{figures/Simulations/181JSCC.png}
	%	\begin{center}
		%	\textbf{\small$0.0612$\\
			%		$20.83/0.807/0.277$}
		%\end{center}
		%\end{minipage}
		%\begin{minipage}[t]{0.07\linewidth}
		%	\centering
		%	\includegraphics[width=1\textwidth]{figures/Simulations/181_Zoomout.png}
		%
		%\end{minipage}
		\hspace{0.3cm}
		\begin{minipage}[t]{0.13\linewidth}
			\centering
			\includegraphics[width=1\textwidth]{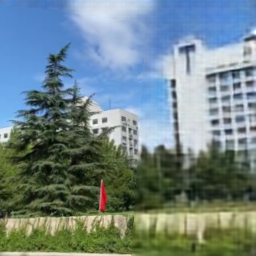}
			\begin{center}
				\textbf{\small$0.0612$\\
					$\textbf{28.14}/\textbf{0.934}/\textbf{0.175}$}
				\vspace*{0.2ex}
			\end{center}
		\end{minipage}

		\begin{minipage}[t]{0.13\linewidth}
			\centering
			\includegraphics[width=1\textwidth]{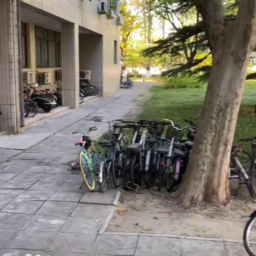}

		\end{minipage}
		\begin{minipage}[t]{0.13\linewidth}
			\centering
			\includegraphics[width=1\textwidth]{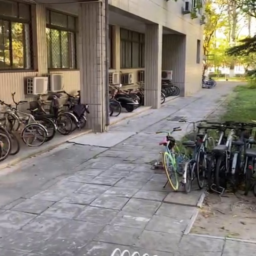}
		\end{minipage}
		\quad
		\begin{minipage}[t]{0.13\linewidth}
			\centering
			\includegraphics[width=1\textwidth]{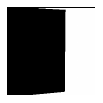}
			\vspace*{-23ex} 
			\vspace*{16ex}
		\end{minipage}
		\quad
		\begin{minipage}[t]{0.13\linewidth}
			\centering
			\includegraphics[width=1\textwidth]{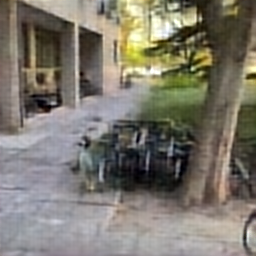}
			\begin{center}
				\textbf{\small$0.0562$\\
					$23.71/0.91$}
			\end{center}
		\end{minipage}
		\quad
		\begin{minipage}[t]{0.13\linewidth}
			\centering
			\includegraphics[width=1\textwidth]{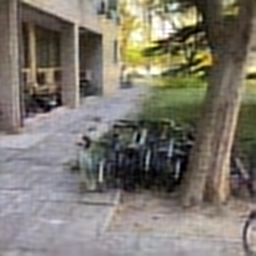}
			\begin{center}
				\textbf{\small$0.0625$\\
					$\textbf{26.26}/0.919/0.26$}
			\end{center}
			
		\end{minipage}
		\quad
		%\begin{minipage}[t]{0.13\linewidth}
		%	\centering
		%	\includegraphics[width=1\textwidth]{figures/Simulations/575JSCC.png}
		%	\begin{center}
			%	\textbf{\small$0.0496$\\
				%		$24.83/0.927/0.147
				%		$}
			%\end{center}
			%\end{minipage}
			\hspace{0.3cm}
			\begin{minipage}[t]{0.13\linewidth}
				\centering
				\includegraphics[width=1\textwidth]{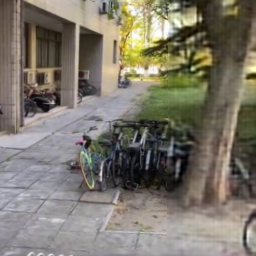}
				\begin{center}
					\textbf{\small$0.0496$\\
						$26.23/\textbf{0.937}/\textbf{0.128}$}
				\end{center}
				\vspace*{0.2ex}
			\end{minipage}

			\begin{minipage}[t]{0.13\linewidth}
				\centering
				\includegraphics[width=1\textwidth]{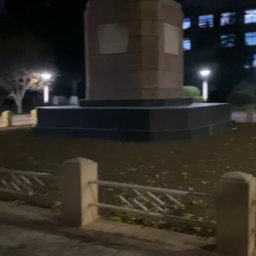}
			\end{minipage}
			\begin{minipage}[t]{0.13\linewidth}
				\centering
				\includegraphics[width=1\textwidth]{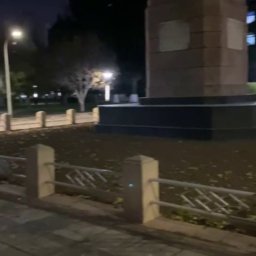}
			\end{minipage}
			\quad
			\begin{minipage}[t]{0.13\linewidth}
				\centering
				\includegraphics[width=1\textwidth]{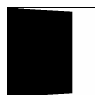}
				\vspace*{-23ex} 
				\vspace*{16ex}
			\end{minipage}
			\quad
			\begin{minipage}[t]{0.13\linewidth}
				\centering
				\includegraphics[width=1\textwidth]{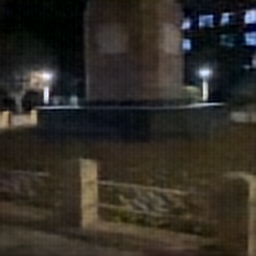}
				\begin{center}
					\textbf{\small$0.0355$\\
						$22.41/0.908$}
				\end{center}
			\end{minipage}
			\quad
			\begin{minipage}[t]{0.13\linewidth}
				\centering
				\includegraphics[width=1\textwidth]{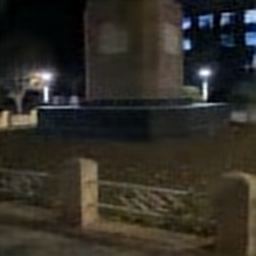}
				\begin{center}
					\textbf{\small$0.03125$\\
						$26.03/0.912/0.29$}
				\end{center}
			\end{minipage}
			\quad
			%\begin{minipage}[t]{0.13\linewidth}
			%	\centering
			%	\includegraphics[width=1\textwidth]{figures/Simulations/800JSCC.png}
			%	\begin{center}
				%	\textbf{\small$0.0317$\\
					%		$26.005/0.952/0.073$}
				%\end{center}
				%\end{minipage}
				\hspace{0.3cm}
				\begin{minipage}[t]{0.13\linewidth}
					\centering
					\includegraphics[width=1\textwidth]{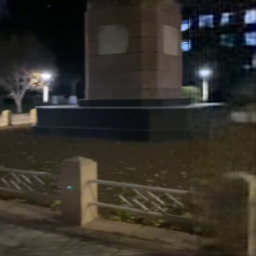}
					\begin{center}
						\textbf{\small$0.0317$\\
							$\textbf{29.30}/\textbf{0.954}/\textbf{0.067}$}
					\end{center}
				\end{minipage}
				\captionsetup{font={small}}
				\caption{Visualization of the comparison on UDIS-D dataset with specific performance at the bottom: the first two row shows the reconstruction on image pair of \textbf{large parallax}; the third row shows the reconstruction on image pair of \textbf{medium parallax}; the last shows the reconstruction on image pair of \textbf{small parallax}. The optimal performance is labeled in bold.}
				\label{Visual1}
			\end{figure*}

To show the robustness of RWZC to varying SSI, we thus consider another dataset consisting of real world samples with varying parallax. Fig. \ref{UDIS_CBR} plots the reconstruct performance of different schemes against CBR on UDIS-D dataset, with fix SNR=3. Unlike the stationary parallax dataset, RWZC shows an obvious advantage on both schemes in terms of three metrics. Specifically, for PSNR, our framework reaches 27.8 dB at CBR$\approx0.031$, while DeepJSCC-WZ only has 26.6 dB and Mital+capacity has 24 dB; for MS-SSIM, RWZC achieves 0.93 while the other two schemes have 0.915 and 0.914 respectively; for LPIPS, the overwhelming lead of our scheme is still exists. In Fig. \ref{UDIS_SNR}, the similar trends against SNR are presented as well.
		\begin{figure*}
			\centering
			\begin{minipage}[t]{0.31\linewidth}
				\centering
				\includegraphics[width=1\textwidth]{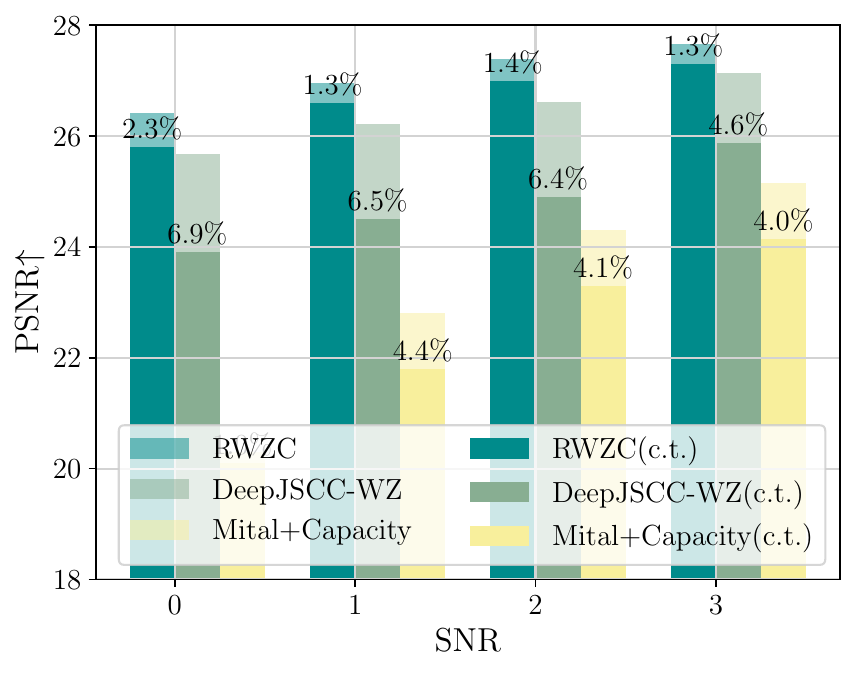}
				%		\vspace*{-37ex} 
				%		\begin{center}
					%			UDIS-D/CBR$\approx$0.031
					%		\end{center}
				%		\vspace*{16ex}
			\end{minipage}
			\begin{minipage}[t]{0.32\linewidth}
				\centering
				\includegraphics[width=1\textwidth]{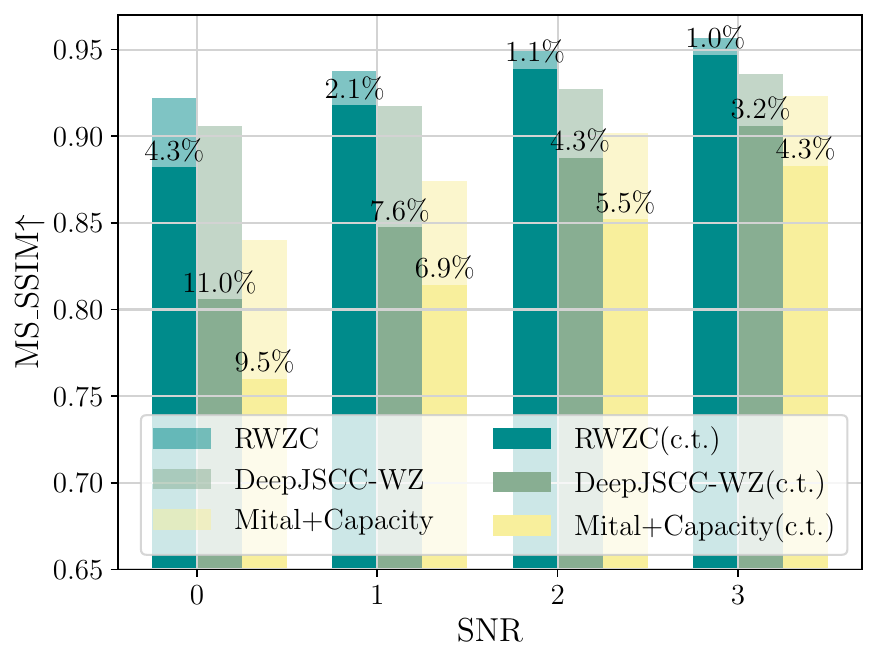}
				%		\vspace*{-37ex} 
				%		\begin{center}
					%			UDIS-D/CBR$\approx$0.031
					%		\end{center}
				%		\vspace*{16ex}
			\end{minipage}
			\begin{minipage}[t]{0.325\linewidth}
				\centering
				\includegraphics[width=1\textwidth]{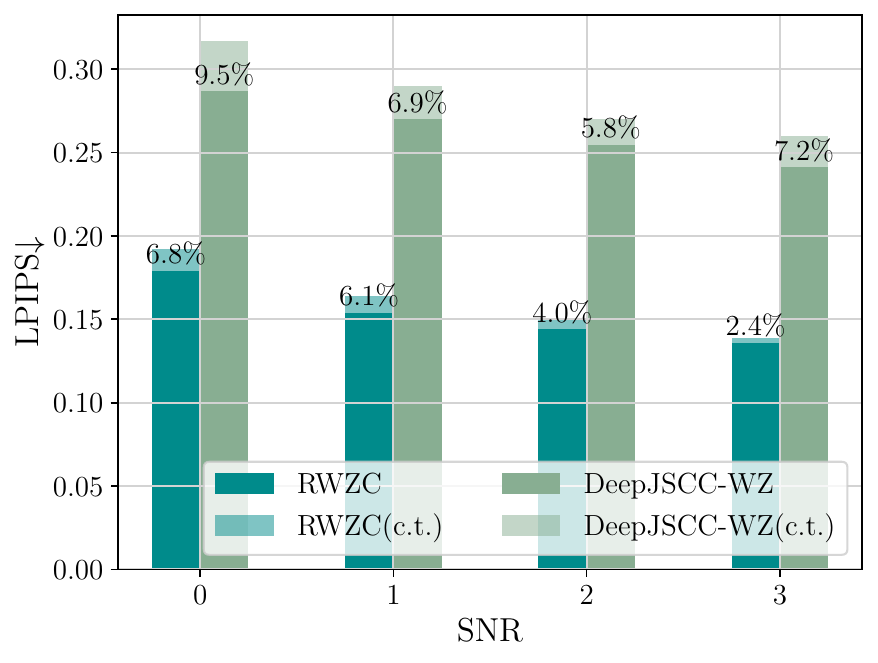}
				%		\vspace*{-37ex} 
				%		\begin{center}
					%			UDIS-D/CBR$\approx$0.031
					%		\end{center}
				%		\vspace*{16ex}
			\end{minipage}
			\captionsetup{font={small},justification=raggedright}
			\caption{Cross-Dataset test of proposed method with the baselines by fixing CBR$\approx0.031$, where herein this test means that the framework is trained on the KITTI stereo and tested on the UDIS-D.}\label{Cross}
		\end{figure*}
	\begin{figure*}[tbp]
		\begin{minipage}[t]{0.49\linewidth}
			\centering
			\includegraphics[width=0.88\textwidth]{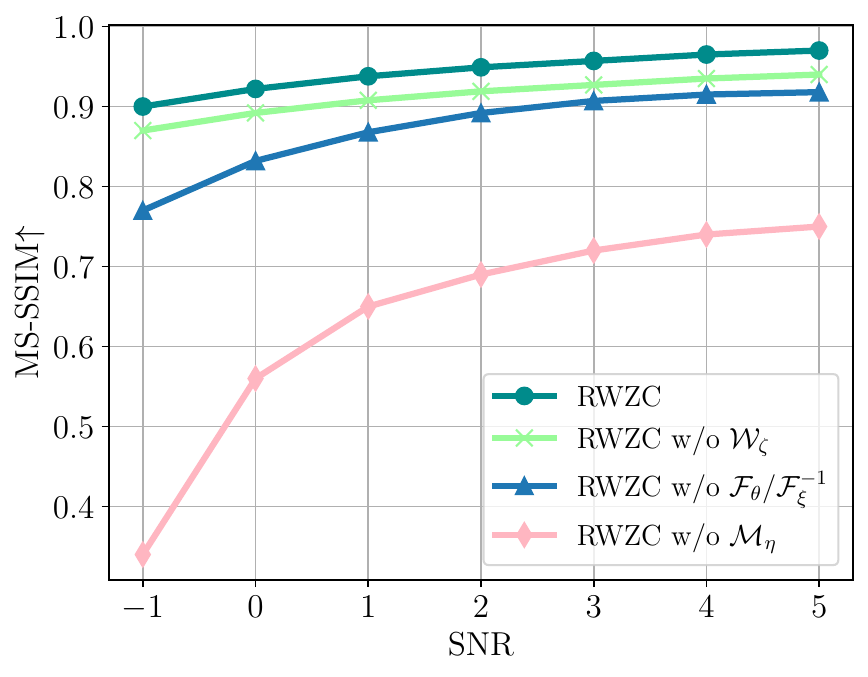}
		\end{minipage}
		\begin{minipage}[t]{0.49\linewidth}
			\centering
			\includegraphics[width=0.9\textwidth]{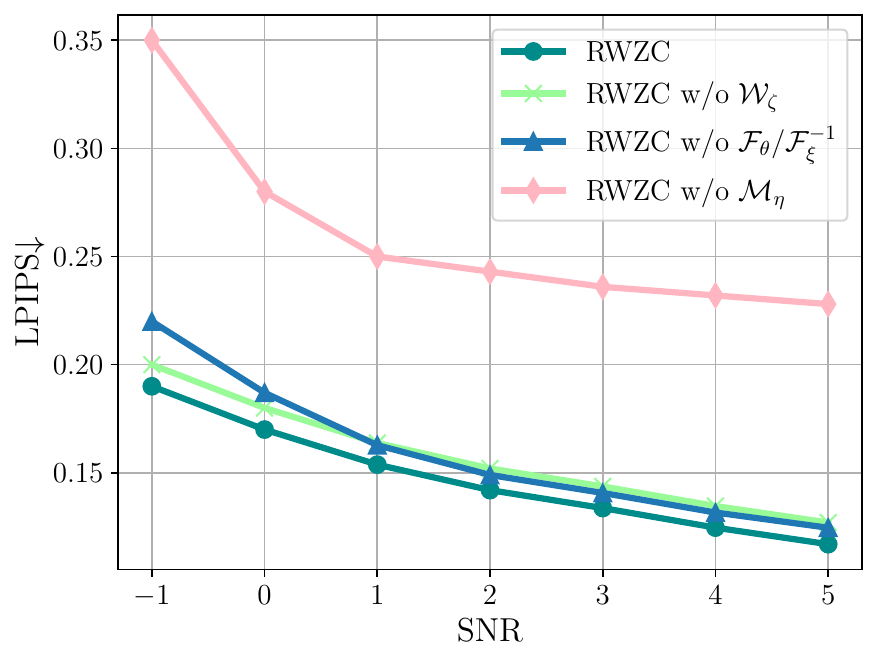}
		\end{minipage}
		\captionsetup{font={small}}
		\caption{Ablation studies on each modules with CBR$\approx$0.031 and SNR=3 dB on UDIS-D dataset.}\label{ablation}
	\end{figure*}

Unfortunately, we find that our scheme is more likely to reach performance saturation (faster convergence in terms of code rate) than the other two, since the warping on pixel level will inevitably brings reconstruction loss, e.g. PSNR and MS-SSIM. Intuitively, RWZC essentially involves performing an affine operation on the original image $\bm{x}$ and then cropping it, which unavoidably leads to the resolution reduction, and thus limits the maximum quality of $\tilde{\bm{y}}$ at the decoder. This means even though we increase the code rate or channel quality, RWZC will reach performance bottleneck earlier than other baselines. Nevertheless, RWZC still offers comparable performance with the state-of-the-art solutions at the low rate interval on stationary-parallax images, while a considerable improvements on real-world images, let alone its outstanding performance under perceptual metrics.
		
Besides, we present some visual results for the comparisons on UDIS-D dataset in Fig. \ref{Visual1} under the same SNR and close CBRs. We select the image pairs from large to small parallaxes, and explore the performance of these frameworks against varying SSI. The first row illustrate the comparison on images of large parallax, which yields a small masked region shown in the column 'Masking(SSI)'. Our proposed scheme outperforms the baselines on all three metrics. The second row also shows the comparison on large parallax image with a totally different SSI, in which RWZC remains ahead and demonstrates it robustness to the varying $\bm{h}$. For the medium parallax (illustrated at the third row), RWZC outperforms others in terms of MS-SSIM and LPIPS, while DeepJSCC-WZ is slightly ahead in PSNR owing to the increasing interactive information at the decoder. 
An interesting finding is that, it is also possible for RWZC outperforms others on specific image pairs of smaller parallax (in the final row). Since the smaller parallax implies more overlapping information, further results in more concealing and lower transmission rate, meanwhile the side information contributes a perfect blending and finally we get a better reconstruction. In a word, it can be verified that our proposed framework can be more effective and suitable for resources-restricted transmission.

\subsection{Cross-dataset Testing}
Moreover, we conduct cross-dataset testing across all baselines. This approach is commonly used in machine learning to evaluate a model's generalization ability, ensuring it performs well on unseen data from potentially different domains, distributions, or environments. Specifically, we fix the CBR around 0.031, train RWZC and the other baselines on the KITTI stereo dataset, and test them on the UDIS-D dataset, where the images are resized to both 370$\times$740.

Figure \ref{Cross} presents the performance comparison among different baselines under cross-dataset testing. We observe that our framework outperforms competitors in terms of all metrics—PSNR, MS-SSIM, and LPIPS—when faced with image samples from unfamiliar distributions. Additionally, it is important to note that cross-dataset testing inevitably leads to some performance degradation. Therefore, we provide a comparison between the in-dataset testing (where both training and testing are done on the same dataset) and cross-dataset testing for each baseline. Moreover, \textbf{the percentage above each bar indicates the performance degradation.} For example, considering the PSNR metric with SNR fixed at 1dB (as shown in the left plot of Figure \ref{Cross}), RWZC with cross-dataset testing achieves 26.59 dB, while the original framework achieves 26.952 dB, representing a $1.3\%$ degradation. Meanwhile, DeepJSCC-WZ shows a degradation of about $6.5\%$, and Mital+Capacity exhibits a $4.4\%$ degradation. We conclude that a smaller percentage indicates less performance degradation, reflecting a stronger robustness when facing unfamiliar data distributions of our framework.

\subsection{Ablation studies}
Recall that our proposed framework consists of three learning-based neural networks, namely masking $\mathcal{M}_{\eta}$, JSC codec $F_{\theta}$ and $F_{\xi}$, and reconstruction $\mathcal{W}_{\zeta}$. In this section we hope to study the respective contributions of these three networks. Before the comparison, we first explain three ablated schemes in the following.

\noindent \textbf{RWZC w/o $\mathcal{M}_{\eta}$} denotes a scheme replaces learning-based masking generation with a fixed masking strategy, and remains the left networks invariant. The fixed masking is realized by an autoencoder only learns the realization of SSI instead of its distribution.

\noindent \textbf{RWZC w/o $F_{\theta}/F_{\xi}$} denotes a scheme replaces the entropy model-based JSC codec with an CNN-based autoencoder of fixed output length 64, and remains the left networks invariant.

\noindent \textbf{RWZC w/o $\mathcal{W}_{\zeta}$} denotes a scheme replaces the warping-prediction recover network with an algorithm-based perspective transform on side image $\bm{y}$ which is mentioned in Sec. \ref{Warp-Pred}, and remains the left networks invariant.

The comparisons among these schemes in terms of MS-SSIM and LPIPS over SNR are shown in Fig. \ref{ablation}. The plots show that the learnable masking strategy contributes the most to the final performance, while the rate-adaptive transmission follows, and the warping prediction reconstruction contributes the least, since the generalization to unfamiliar SSI is essential to the framework. The trends turn a slight different in LPIPS comparison, where the warping prediction module hardly influences the performance since the shifting of a few pixels will not affect the human perception.

\begin{table}[b]
	\centering
	\caption{Comparisons on time and storage complexities}\label{complexity}
	\begin{tabular}{m{1cm}<{\centering}|m{1.5cm}<{\centering}|m{2cm}<{\centering}|m{1.5cm}<{\centering}}
		\hline
		\rowcolor[gray]{0.9} % 设置行的背景色
		\centering &  RWZC &\centering Mital (Only compression)& DeepJSCC-WZ \\
		\hline
		%\rowcolor[gray]{0.8} % 设置行的背景色
		\centering FLOPs (G) & 58.72 & 60.06& -\\
		\hline
		\centering Param (M)& 26.19 & 24.44& 48.9 \\
		\hline
	\end{tabular}
\end{table}
\vspace{0.5cm}
\subsection{Complexity analysis}

We refer to Table \ref{complexity} for comparisons on computation and spatial complexities. We remark that RWZC enjoys a moderate complexity involving learnable masking and warping networks. For comparison, The scheme from Mital conducts the feature extraction of common information at the decoder hence has a slightly higher FLOPs, in which we only consider the compression. Meanwhile, DeepJSCC-WZ utilizing the feature attention module will suffer from relative higher computation and storage burdens. This comparison verifies the complexity degradation from a data-driven to a proper model-driven learning-based framework.

\section{Conclusion}\label{Sec6}
In this work, we propose a novel learning-based Wyner-Ziv coding framework that is robust to non-stationarily correlated sources. Unlike existing data-driven approaches, our framework establishes an affine model between images, effectively separating the overlapping information of the transmitted image from the side image, enabling reliable and efficient communication. These schemes have been shown to outperform state-of-the-art competitors, particularly in terms of perceptual metrics.

For future work, we plan to explore different modeling approaches for source correlation in the RWZC framework, which may involve non-linear correlations, such as those arising from illuminations and depth, rather than just affine correlation. Non-linear correlation modeling can better align with real-world sources but require more advanced network architecture design. Additionally, we aim to extend our framework to better accommodate varying channel environments, including fading channels and MIMO systems.

% \section*{Acknowledgments}
% This should be a simple paragraph before the References to thank those individuals and institutions who have supported your work on this article.

\small
\bibliographystyle{IEEEtran}
\bibliography{bib1571028021}
\end{document}